\documentclass[twocolumn]{aastex63}
\usepackage{amsmath}
\usepackage{mathrsfs}
\usepackage{xcolor}
\usepackage[T1]{fontenc}
\usepackage{fourier}
\usepackage{hyperref}
\usepackage[normalem]{ulem}
\usepackage{relsize}
\usepackage{graphicx}
\usepackage{subfigure}
\submitjournal{ApJ}

\shorttitle{Proposed CHIME/FRB Host Galaxies}
\shortauthors{Ibik et al.}

\begin{document}
  
\title{Proposed host galaxies of repeating fast radio burst sources detected by CHIME/FRB}

\correspondingauthor{Adaeze Ibik}
\email{adaeze.ibik@mail.utoronto.ca}

\author[0000-0003-2405-2967]{Adaeze L.~Ibik}
  \affiliation{David A.~Dunlap Department of Astronomy \& Astrophysics, University of Toronto, 50 St.~George Street, Toronto, ON M5S 3H4, Canada}
  \affiliation{Dunlap Institute for Astronomy \& Astrophysics, University of Toronto, 50 St.~George Street, Toronto, ON M5S 3H4, Canada}

\author[0000-0001-7081-0082]{Maria R. Drout}
  \affiliation{David A.~Dunlap Department of Astronomy \& Astrophysics, University of Toronto, 50 St.~George Street, Toronto, ON M5S 3H4, Canada}
  \affiliation{Dunlap Institute for Astronomy \& Astrophysics, University of Toronto, 50 St.~George Street, Toronto, ON M5S 3H4, Canada}

  \author[0000-0002-3382-9558]{B.~M.~Gaensler}
  \affiliation{Dunlap Institute for Astronomy \& Astrophysics, University of Toronto, 50 St.~George Street, Toronto, ON M5S 3H4, Canada}
  \affiliation{David A.~Dunlap Department of Astronomy \& Astrophysics, University of Toronto, 50 St.~George Street, Toronto, ON M5S 3H4, Canada}
  \affiliation{Present address: Division of Physical and Biological Sciences, University of California Santa Cruz, 1156 High Street, Santa Cruz, CA 95064, USA}
  \author[0000-0002-7374-7119]{Paul Scholz}
  \affiliation{Dunlap Institute for Astronomy \& Astrophysics, University of Toronto, 50 St.~George Street, Toronto, ON M5S 3H4, Canada}
  \author[0000-0002-2551-7554]{Daniele Michilli}
  \affiliation{MIT Kavli Institute for Astrophysics and Space Research, Massachusetts Institute of Technology, 77 Massachusetts Ave, Cambridge, MA 02139, USA}
  \affiliation{Department of Physics, Massachusetts Institute of Technology, 77 Massachusetts Ave, Cambridge, MA 02139, USA}
\author[0000-0002-3615-3514]{Mohit Bhardwaj}
  \affiliation{Department of Physics, Carnegie Mellon University, 5000 Forbes Avenue, Pittsburgh, 15213, PA, USA}
\author[0000-0001-9345-0307]{Victoria M.~Kaspi}
  \affiliation{Department of Physics, McGill University, 3600 rue University, Montr\'eal, QC H3A 2T8, Canada}
  \affiliation{Trottier Space Institute, McGill University, 3550 rue University, Montr\'eal, QC H3A 2A7, Canada}
\author[0000-0002-4795-697X]{Ziggy Pleunis}
  \affiliation{Dunlap Institute for Astronomy \& Astrophysics, University of Toronto, 50 St.~George Street, Toronto, ON M5S 3H4, Canada}

\author[0000-0003-2047-5276]{Tomas Cassanelli}
  \affiliation{Department of Electrical Engineering, Universidad de Chile, Av. Tupper 2007, Santiago 8370451, Chile}
\author[0000-0001-6422-8125]{Amanda M. Cook}
  \affiliation{David A.~Dunlap Department of Astronomy \& Astrophysics, University of Toronto, 50 St.~George Street, Toronto, ON M5S 3H4, Canada}
  \affiliation{Dunlap Institute for Astronomy \& Astrophysics, University of Toronto, 50 St.~George Street, Toronto, ON M5S 3H4, Canada}
\author[0000-0003-4098-5222]{Fengqiu A. Dong}
  \affiliation{Department of Physics and Astronomy, University of British Columbia, 6224 Agricultural Road, Vancouver, BC V6T 1Z1 Canada}
 \author[0000-0003-4810-7803]{Jane F.~Kaczmarek}
\affiliation{Dominion Radio Astrophysical Observatory, Herzberg Research Centre for Astronomy and Astrophysics, National Research Council Canada, PO Box 248, Penticton, BC V2A 6J9, Canada}
\affiliation{CSIRO Astronomy and Space Science, Parkes Observatory, P.O. Box 276, Parkes NSW 2870, Australia}
\author[0000-0002-4209-7408]{Calvin Leung}
  \affiliation{MIT Kavli Institute for Astrophysics and Space Research, Massachusetts Institute of Technology, 77 Massachusetts Ave, Cambridge, MA 02139, USA}
  \affiliation{Department of Physics, Massachusetts Institute of Technology, 77 Massachusetts Ave, Cambridge, MA 02139, USA}

\author{Katherine J.~Lu}
   \affiliation{Department of Physics, Carnegie Mellon University, 5000 Forbes Avenue, Pittsburgh, 15213, PA, USA}
\author[0000-0002-4279-6946]{Kiyoshi W.~Masui}
  \affiliation{MIT Kavli Institute for Astrophysics and Space Research, Massachusetts Institute of Technology, 77 Massachusetts Ave, Cambridge, MA 02139, USA}
  \affiliation{Department of Physics, Massachusetts Institute of Technology, 77 Massachusetts Ave, Cambridge, MA 02139, USA}
\author[0000-0002-8912-0732]{Aaron B.~Pearlman}
  \affiliation{Department of Physics, McGill University, 3600 rue University, Montr\'eal, QC H3A 2T8, Canada}
  \affiliation{Trottier Space Institute, McGill University, 3550 rue University, Montr\'eal, QC H3A 2A7, Canada}
\author[0000-0001-7694-6650]{Masoud Rafiei-Ravandi}
  \affiliation{Department of Physics, McGill University, 3600 rue University, Montr\'eal, QC H3A 2T8, Canada}
  \affiliation{Trottier Space Institute, McGill University, 3550 rue University, Montr\'eal, QC H3A 2A7, Canada}
\author[0000-0003-3154-3676]{Ketan R Sand}
  \affiliation{Department of Physics, McGill University, 3600 rue University, Montr\'eal, QC H3A 2T8, Canada}
  \affiliation{Trottier Space Institute, McGill University, 3550 rue University, Montr\'eal, QC H3A 2A7, Canada}
  \author[0000-0002-6823-2073]{Kaitlyn ~Shin}
  \affiliation{MIT Kavli Institute for Astrophysics and Space Research, Massachusetts Institute of Technology, 77 Massachusetts Ave, Cambridge, MA 02139, USA}
  \affiliation{Department of Physics, Massachusetts Institute of Technology, 77 Massachusetts Ave, Cambridge, MA 02139, USA}
\author[0000-0002-2088-3125]{Kendrick M.~Smith}
  \affiliation{Perimeter Institute for Theoretical Physics, 31 Caroline Street N, Waterloo, ON N25 2YL, Canada}
\author[0000-0001-9784-8670]{Ingrid H.~Stairs}
  \affiliation{Department of Physics and Astronomy, University of British Columbia, 6224 Agricultural Road, Vancouver, BC V6T 1Z1 Canada}

\begin{abstract}
We present a search for host galaxy associations for the third set of repeating fast radio burst (FRB) sources discovered by the CHIME/FRB Collaboration. Using the $\sim$ 1 arcmin CHIME/FRB baseband localizations and probabilistic methods, we identify potential host galaxies of two FRBs, 20200223B and 20190110C at redshifts of 0.06024(2) and 0.12244(6), respectively. We also discuss the properties of a third marginal candidate host galaxy association for FRB 20191106C with a host redshift of 0.10775(1).  The three putative host galaxies are all relatively massive, fall on the standard mass-metallicity relationship for nearby galaxies, and show evidence of ongoing star formation. They also all show signatures of being in a transitional regime, falling in the ``green valley'' which is between the bulk of star-forming and quiescent galaxies. The plausible host galaxies identified by our analysis are consistent with the overall population of repeating and non-repeating FRB hosts while increasing the fraction of massive and bright galaxies. Coupled with these previous host associations, we identify a possible excess of FRB repeaters whose host galaxies have $M_{\mathrm{u}}-M_{\mathrm{r}}$ colors redder than the bulk of star-forming galaxies. Additional precise localizations are required to confirm this trend. 

\end{abstract}


\section{Introduction} \label{sec:intro}

Over 600 short-duration energetic radio pulses known as fast radio bursts \citep[FRBs;][]{Lorimer2007} have been reported\footnote{\url{https://www.chime-frb.ca/catalog}} \citep{chime-amiri2021, Nimmo2023}. Multiple FRBs that are coincident in sky location and dispersion measure (DM)---``FRB repeaters'', are particularly interesting since they provide more opportunities for detailed follow-up studies.  While we do not yet know whether repeaters and one-off FRBs are two separate classes, the existence of repeaters demonstrates that the progenitors of at least some FRBs are not cataclysmic (for a review of FRB models, see \citealt{Platts_2019}). Recently, the magnetar theory for FRBs \citep{Beloborodov2017, metzger2019} was confirmed as one possible origin of FRB with the discovery of a very luminous FRB-like burst from the Galactic magnetar SGR $1935 + 2154$ \citep{CHIME-Anderson2020,Bochenek2020}. However, it remains the only FRB source confirmed to be a magnetar, emphasizing the need for additional probe of their nature.

The host galaxies of various types of transients have often provided useful insight into their origins---allowing constraints on the ages and physical properties required to produce them \citep[e.g.,][]{Eftekhari2019, Eftekhari2021}.  To date, approximately 34 FRBs have been localized to their host galaxies \citep[e.g.,][]{Chatterjee_2017, Bhandari2020,Heintz2020,Bhandari2021, BhardwajR42021,Niu2021, Bhandari2022,Michilli2022,Rajwade2022meerkat,Gordon2023,Sharma2023,Jankowski2023}. 
This is due to the difficulty of pinpointing FRBs within the arcsecond precision required for robust host associations \citep{EftekhariandBerger2017} and multi-wavelength follow-up \citep[e.g.,][]{Scholz2017}.  
The host populations of both repeating and non-repeating FRBs have diverse properties (star-forming/quiescent, low/high metallicity, massive/dwarf, spiral/elliptical) \citep[e.g.,][]{Heintz2020, Bhandari2021,bha+22} and it is not clear what type of host is preferred by repeaters. 
While the nature and the origin of repeating FRBs are still debated, finding more host galaxies will provide insights into the origin of the FRB population. In addition to understanding the nature of FRBs, FRBs with known host galaxies are useful cosmological probes because the relationship between their DMs and redshifts could give insight into the missing baryons in the intergalactic medium (IGM) \citep{Macquart_2020}.

Recently, the Canadian Hydrogen Intensity Mapping Experiment Fast Radio Burst (CHIME/FRB) project \citep{CHIME2018} published 25 new repeaters and a further 14 candidates as described by \cite{CHIME2023}. This supplements
 21 previously published CHIME/FRB repeaters \citep{2019Natur.566..235C,CHIME2019b,Fonseca2020, Bhardwaj_2021,Lanman2022,McKinven2022R117}, all of which are observed in the frequency range 400–800\,MHz. This sample represents by far the largest population of repeaters detected by a single survey to date and thus offers the opportunity to learn about various properties of the FRB population. CHIME/FRB repeaters show a lower average dispersion measure than those of apparently non-repeating sources \citep{CHIME2023} even though this could be a selection effect \citep[see][]{Connor2020}.
 
Other previously established characteristic properties of repeaters such as wider bursts and narrower emitting bandwidths than non-repeating FRBs were confirmed in the additional set of repeaters \citep{chime-amiri2021,Pleunis2021}. These observational differences could signal a distinct astrophysical origin for repeaters. Additional multi-wavelength studies with a larger sample, including host galaxy identification, could help to disentangle potential multiple populations.

Typically, robust host associations require  $\lesssim$ 1\,arcsecond localizations for the FRBs, and hence detection of a burst with interferometers \citep{EftekhariandBerger2017}. Indeed, some FRB repeaters (e.g., FRB 20180916B and 20201124A; \citealt{Marcote2020}, \citealt{Nimmo2022}) discovered by CHIME/FRB have subsequently been associated with their host galaxies by means of very long-baseline interferometry (VLBI). However, arcsecond localizations are not available for most CHIME/FRB bursts. At best, CHIME/FRB can achieve $\sim$1\,arcminute localization for bursts when baseband data are recorded. CHIME/FRB baseband data are channelized raw voltages of the FRB signals from the telescope feeds. Localizations with precision $\sim$1\,arcminute using baseband data from CHIME/FRB can be obtained following the techniques described by \cite{Michilli_2021}. While these baseband positions are larger than the size of a typical galaxy in the sky, for objects in the relatively nearby universe, it is possible to make probable host galaxy associations due to the low density of galaxies at lower redshifts. This technique has previously been used to associate five FRBs detected by CHIME/FRB with likely hosts: FRBs 20200120E \citep{Bhardwaj_2021}, 20181030A \citep{BhardwajR42021}, 20220912A \citep{McKinven2022R117}, 20180814A, and 20190303A \citep{Michilli2022} and the reliability of the method was supported by the eventual VLBI confirmation of the host galaxies for FRB 20200120E \citep{kirsten2021}. 

Future telescopes such as the CHIME/FRB Outrigger project (currently under construction) will be able to improve the number of precisely localized FRBs \citep{Leung2021,Mena-Parra2022,Cassanelli2022}. Other such facilities include the Real-time fast transients at the Karl Janskly Very Large Array \citep[realfast VLA;][]{Law2018}, the fast radio transient-detection program at MeerKAT \citep[MeerTRAP;][]{Rajwade2022meerkat}, and the Deep Synoptic Array \citep[DSA-110;][]{Ravi2022DS110}. Until sub-arcseconds localization is possible for CHIME/FRB events, the baseband position remains a useful technique for associating local universe FRBs to their host galaxies \citep{Michilli2022}. 

In this paper, we perform a search for likely host galaxies for 20 repeaters and 3 repeater candidates from the recently published sample by \cite{CHIME2023} for which $\sim$arcmin baseband localization were available. In total, we identify highly probable hosts for two FRBs and one marginally significant candidate host for a third. A thorough description of our search for the probable hosts of the events is described in \S \ref{sec:sample-selection} and details of the observations of likely hosts are explained in \S \ref{sec:observations-likely-host}. The observed and inferred properties of the hosts are described in \S \ref{sec:properties-of-host} including a comparison between the three hosts and previously known hosts of FRBs in \S \ref{subsec:comparison-to-other-FRBs}. Finally, we summarize the result of the host associations in \S \ref{sec:summary}.

\section{Search for Probable Host Galaxies} \label{sec:sample-selection}

Here we describe the process used to search for probable host galaxies to the CHIME/FRB repeaters. Specifically, after describing our initial sample (\S \ref{subsec:initial-sample}), we determine the maximum redshift for each FRB based on its DM (\S \ref{subsec:Determination-of-zmax}), search archival optical catalogs for galaxies within the localization regions for each burst (\S \ref{subsec:optical-catalogs}), and perform a probability of chance coincidence analysis (\S \ref{subsec:Pcc}). Finally, we summarize these results for three galaxies identified as potential hosts to CHIME repeating FRBs in \S \ref{subsec:summary-robust-host}.

\subsection{Initial FRB Sample} \label{subsec:initial-sample}
Of the 25 `gold' and 14 `silver' repeating FRBs presented by \cite{CHIME2023}, we first restrict ourselves to objects that have baseband localizations for multiple bursts because of their small localization uncertainties. This criterion is met for 20 gold and 3 silver candidates.  From there, we continue the analysis with events that are in the survey coverage area of either the Sloan Digital Sky Survey (SDSS) DR12 \citep{sdss2015} or Dark Energy Spectroscopy Instrument (DESI), Data Release 9 (DR9) \citep{dsl+19} Legacy Imaging Survey photometric catalog, as these are required to assess the population of possible host galaxies within the CHIME/FRB localizations regions. This reduced the sample list to the 12 gold and 1 silver (FRB 20190303D) candidates listed in Table \ref{tab:catalogue details}.

\begin{deluxetable*}{l|ccc|ccc|ccc}
\tabletypesize{\small}
\tablecaption{Summary of FRB repeater candidates searched for likely host galaxies \label{tab:catalogue details}}
\tablehead{\colhead{FRB Name} & \colhead{DM} & \colhead{DM$_\mathrm{MW}$$^{c}$} & \colhead{z$_{\mathrm{max}}$} & \colhead{$N_{\mathrm{s}}$$^{d}$} 
& \colhead{Max $P_{\mathrm{\texttt{PATH}}}$$^{e}$} &  \colhead{Max $P_{\mathrm{\texttt{PATH}}}$$^{e}$} &  \colhead{m$_{\mathrm{r,\mathrm{gal}}}$$^{f}$} &  \colhead{$P_\mathrm{cc,gal}$$^{g}$} \\
\colhead{} & \colhead{(pc\,cm$^{-3}$)} & \colhead{(pc\,cm$^{-3}$)} & \colhead{} & \colhead{} 
& \colhead{$\mathrm{P(U)} =0.0$} & \colhead{$\mathrm{P(U)} =0.1$} & \colhead{(AB mag)} & \colhead{}}
\startdata
     FRB 20200223B & 201.8(4) & 45.6 & 0.19 & 7 & 0.994 & 0.899 & 16.08 & 0.010 \\
      FRB 20191106C & 332.2(7) & 25.0 & 0.36 & 11 
      & 0.951 & 0.815 & 17.31 & 0.063 \\ 
       FRB 20190110C & 221.6(1.6) & 37.1 & 0.22 & 9  
      & 0.918 & 0.779 & 18.00 & 0.102 \\
       FRB 20190804E & 363.2(3) & 43.4 & 0.37 & 9 
         & 0.638 & 0.549 & 19.13  & 0.305 \\
        FRB 20200118D & 625.7(1.4) & 76.6 & 0.62 & 14 
        & 0.619 & 0.520 & 19.71 & 0.491 \\
         FRB 20201114A & 321.1(8) & 37.8 & 0.33 & 60 
         & 0.537 & 0.454 & 17.68 & 0.469\\ 
               FRB 20200913C & 574.2(1.6) & 47.8 & 0.60 & 74
      &  0.490 & 0.407 & 17.87 & 0.600\\
     FRB 20200929C & 413.3(3) & 38.4 & 0.44 & 41 
     & 0.474 & 0.388 & 17.04 & 0.144 \\
      FRB 20191013D & 524.1(6.5) & 43.1 & 0.55 & 111
    & 0.465 & 0.396 & 17.49 & 0.401 \\
     FRB 20190609C$^{a}$ & 479.2(3) & 113.4 & 0.43 & 3
          & 0.409 & 0.354 & 22.58 & 0.745\\
     FRB 20201221B & 509.5(1.1) & 50.9 & 0.53 & 45  
     & 0.210 & 0.175 & 19.48 & 0.686 \\
      FRB 20190303D$^{b}$ & 716.6(1.6) & 36.5 & 0.77 & 28
      & 0.118 & 0.098 & 20.55 & 0.750\\
       FRB 20200619A & 439.6(0.4) & 50.8 & 0.45 & 350 
       & 0.080 & 0.067 & 18.18 & 0.753 \\
\enddata
\textbf{Notes:} \\
$^{a}$ \texttt{PATH} probability using the SDSS list of host candidates while others used the DESI list of host candidates. \\
$^{b}$ This is the only silver candidate in our sample. The rest of the candidates are from the gold sample.\\
$^{c}$ DM contribution expected for the disk of the Milky Way from the NE2001 model \citep{ne2001}. \\
$^{d}$ $N_{\mathrm{s}}$ is the number of extended optical objects detected in the FRB 90\% error region. \\
$^{e}$ Max $P_{\mathrm{\texttt{PATH}}}$ is the maximum probability estimated by \texttt{\texttt{PATH}} for each FRB region given two different priors, $\mathrm{P(U)} =0.0$ and $\mathrm{P(U)} =0.1$.  FRBs are ordered in the decreasing order of Max $P_{\mathrm{\texttt{PATH}}}$.\\
$^{f}$ m$_{\mathrm{r, \mathrm{gal}}}$(AB) is the r-band AB magnitude of the plausible host of the FRB as identified by \texttt{\texttt{PATH}}. \\
$^{g}$ $P_\mathrm{cc,gal}$ is the probability of finding a galaxy with the r-band magnitude of the most probable host or brighter ($<$m$_{\mathrm{r, \mathrm{gal}}}$) within each FRB 90\% error region.  \\
\end{deluxetable*}

\subsection{Determination of z$_{\mathrm{max}}$} \label{subsec:Determination-of-zmax}

To begin, we calculate the maximum redshift for each FRB in our sample. This is possible because the DM measured for each burst gives us an integrated column density of electrons along the line of sight. This DM will contain contributions from materials in the Milky Way (MW) disk and halo, the inter-galactic medium (IGM), the host galaxy interstellar medium (ISM), and potentially the local material around the FRB. However, if electrons in the IGM are the dominant contribution, the DM of FRBs is approximately linearly related to redshift.

For each of the FRBs in Table \ref{tab:catalogue details}, we determine the maximum redshift $z_\mathrm{max}$ using DMs taken from \cite{CHIME2023}. To do this, we subtract the DM contribution expected for the disk of the Milky Way from the NE2001 model \citep{ne2001}, DM$_\mathrm{MW}$(NE2001), but do not attempt to correct for contributions from either the MW halo (DM$_\mathrm{halo}$) or the FRB host galaxy (DM$_\mathrm{host}$). 

This method provides us with a conservative estimate for the ``excess'' DM associated with each burst (DM$_\mathrm{excess}$). We then calculate the maximum redshift, z$_{\mathrm{max}}$ for each repeater event by attributing this entire DM$_\mathrm{excess}$ to the IGM using the \texttt{FRUITBAT} software \citep{Batten2019}. \texttt{FRUITBAT} uses the given DM to estimate the maximum redshift given the cosmology of \cite{Komatsu2009} and the DM-redshift relationship of \cite{Zhang2018}. In Table \ref{tab:catalogue details}, we list the DM, DM$_\mathrm{MW}$(NE2001), and z$_{\mathrm{max}}$ for all bursts in our sample. The derived z$_{\mathrm{max}}$ values span 0.19 $<$ z$_{\mathrm{max}}$ $<$ 0.77. 

We acknowledge that this method is different from that described by \cite{Bhardwaj_2021} and used in other CHIME/FRB papers. However, we emphasize that z$_{\mathrm{max}}$ is not utilized in the probabilistic analysis below. We consider all galaxies in the field of the FRB. Rather, z$_{\mathrm{max}}$ is only used to ensure that any individual host with a high probability of association has z < z$_{\mathrm{max}}$ (see Section~\ref{subsec:summary-robust-host}). However, using the method described by \cite{Bhardwaj_2021} does not change the set of possible host galaxies. We, therefore, opt for the most conservative approach.

\subsection{Optical Galaxy Catalog Search} \label{subsec:optical-catalogs}
We searched the SDSS and DESI catalogs for host galaxy candidates within the 90\%-confidence baseband positions of the FRBs. For 12 of the 13 events that are in the survey coverage area of both DESI and SDSS, we adopted the DESI catalog which is more sensitive to fainter sources (r-band 5$\sigma$ depths of $\sim$23.4 mag), while the SDSS catalog (r-band 5$\sigma$ depths of $\sim$22.7 mag) was used to analyze FRB 20190609C. The impact of completeness in these catalogs, especially for faint galaxies, will be discussed below. To identify likely host galaxies, we first exclude any object classified as a point source within the SDSS or DESI catalogs. Multiple extended sources were found within the 90\%-confidence positional error region for each FRB.  
We record the number of extended objects within the localization regions for each FRB as $N_\mathrm{s}$ in Table \ref{tab:catalogue details}.

\subsection{Host Probability Analysis} \label{subsec:Pcc}

We use two complementary methods to assess the likelihood that a galaxy within the CHIME/FRB error region can be robustly associated with a host galaxy. We use the result of these methods together to decide on when/whether likely host associations can be made. 

\emph{First method: } First, for cases where more than one galaxy was identified in DESI/SDSS imaging, we analyze the probability of each being associated with the FRB using the formalism described in the Probabilistic Association of Transients to their Hosts (\texttt{\texttt{PATH}}) software \citep{Aggarwal2021}. \texttt{PATH} takes as input the FRB position and its uncertainty, and the r-band magnitudes and half-light radii of all candidate galaxies. It then calculates a probability for each galaxy using Bayes' theorem and accounting for the FRB localization error region. We initially adopt priors that assume (i) the probability that each candidate galaxy is the host and is inversely proportional to the angular surface density of galaxies with magnitudes equal to or brighter than the candidate's magnitude (based on galaxy counts from \citealt{Driver2016}), (ii) the probability of finding the host galaxy of the FRB at a given physical offset is an exponentially declining function, and (iii) that the probability that the true host was undetected, $\mathrm{P(U)} $, was zero ($\mathrm{P(U)} = 0$). The impact of adjusting this final assumption on the resultant probabilities is examined below, as well as how reasonable the assumption is given the depth of the DESI/SDSS surveys at the maximum redshifts of specific bursts (see \S ~\ref{subsec:summary-robust-host}). For the CHIME/FRB host candidates described above, we obtained the half-light radii in arcseconds from DESI and used the Petrosian radius instead in the absence of a half-light radius for the FRB with only SDSS data. 

Applying the \texttt{PATH} framework, we obtain a probability of associating the FRB with each of the candidates identified in its localization region and report the highest probability value, Max $P_{\mathrm{\texttt{PATH}}}$ in Table \ref{tab:catalogue details}. 
In total, we find three FRBs where the maximum \texttt{PATH} probability ($P_{\mathrm{\texttt{PATH}}}$) for the $\mathrm{P(U)} = 0$ case is $>$ 90\%. Given that we initially ran \texttt{PATH} adopting a probability of zero that the true host was undetected (and hence not in our candidate list), this is effectively a statement that \texttt{PATH} significantly prefers one of the candidate hosts over all \emph{detected} galaxies\footnote{The resulting \texttt{PATH} probabilities for all galaxies within these fields are given in Appendix \ref{appendix}}. For the remaining 10 FRBs, \texttt{PATH} gives a maximum probability to any individual galaxy of  $ \lesssim$ 75\% \emph{even} in the case that the probability of an undetected host is zero. For these cases, we argue that it is not possible to make a likely host association with current data, while the three FRBs with higher Max $P_{\mathrm{\texttt{PATH}}}$ warrant further examination.

In particular, it is possible that some potential host galaxies were missing from our initial candidate lists. The DESI detection limit corresponds to an absolute r-band magnitude of $-$15.9 to $-$17.48 at redshifts between 0.19 and 0.36 (the range of $z_{\mathrm{max}}$ values for the three bursts with the highest \texttt{PATH} probability hosts). At these depths, we are sensitive to galaxies fainter than 0.1L$_*$ including all of the superluminous supernova and long-duration gamma-ray burst host galaxies at $z<$0.3 of \cite{Lunnan2014}. However, some of the faintest dwarf galaxies, such as the M$_r$ $>$ $-$16.5 host of FRB121102 \citep{Chatterjee_2017}, could have eluded detection at the high-end of this distance range. 

Given the possibility of undetected dwarf galaxies, we re-run \texttt{PATH}, now setting an arbitrary prior on the probability that the true host was undetected as $\mathrm{P(U)} = 0.1$ (recommended by the authors of \texttt{PATH}). The results are also listed in Table \ref{tab:catalogue details}. For one FRB, 20200223B, the galaxy with the maximum \texttt{PATH} probability is approximately 90\%, while for the other two (20190110C, 20191106C), the maximum \texttt{PATH} probability falls to 82\% $-$ 78\%. We will further discuss the appropriateness of these values of P(U) for each of these specific FRBs---given our survey depth at their maximum distance---in Section~\ref{subsec:summary-robust-host}.

\emph{Second method: } Given that the appropriate value for $\mathrm{P(U)} $ for each burst is uncertain, we also calculate the probability that the galaxy \emph{most preferred by \texttt{PATH}} would be found within the CHIME localization region for that FRB by chance. To calculate this probability of chance coincidence ($P_\mathrm{cc,gal}$), we use a Monte Carlo approach. Specifically, for each FRB, we query random patches of the sky within the SDSS footprint having the size of the FRB localization region for 1000 times. We used SDSS instead of DESI because it was easier to query the entire SDSS catalog multiple times for many error regions. We determine the number of galaxies within that patch that have r-band magnitudes brighter than or equal to that of the most probable host galaxy according to \texttt{PATH} (m$_{\mathrm{r}}$ $\leq$ m$_{\mathrm{r,\mathrm{gal}}}$). We then use the fractions of these runs that have at least one galaxy that meets this requirement to determine the probability that the most plausible identified host was within the FRB 90\%-confidence localization region by chance. In Table \ref{tab:catalogue details}, we list $P_\mathrm{cc,gal}$ as well as the AB r-band magnitude (m$_{\mathrm{r,\mathrm{gal}}}$) values for the preferred host of these FRBs.

We note that these values are most useful in cases where the maximum \texttt{PATH} probability is high. For example, while the galaxy with the highest \texttt{PATH} probability within the localization region of FRB 20200929C has a small $P_\mathrm{cc,gal}$ of 14.4\%, \texttt{PATH} only assigns a probability of $\sim$38 $-$ 47\% that this is the true host because there are two galaxies of similar brightness within the localization region. On the other hand, for three FRBs (20200223B, 20190110C, 20191106C) with maximum \texttt{PATH} probability at $\mathrm{P(U)} = 0$ $ > $90\%, we calculate that there was a $ \leq$10\% chance that the preferred galaxy is located within the CHIME/FRB localization region by chance. 

We emphasize that these $P_\mathrm{cc,gal}$ values describe the probability that an \emph{individual} galaxy would be located within an area the size of an \emph{individual} FRB localization region. They do not account for ``look-elsewhere'' effects associated with the fact that we have searched multiple CHIME/FRB localization regions as part of this overall study. If we instead perform our Monte Carlo analysis considering a region with the \emph{combined} area of all 13 FRBs listed in Table~\ref{tab:catalogue details}, we find that: (i) in 71.7\% of trials, there is at least one galaxy with $m_r < 16.28$ (the SDSS magnitude of the preferred host for FRB 20200223B), (ii) in 72.9\% of trials, there are at least 3 galaxies with $m_r < 17.43$ (the SDSS magnitude of the preferred host for FRB 20191106C), and (iii) in 73.6\% of trials, there are at least 8 galaxies with $m_r < 18.18$  (the SDSS magnitude of the preferred host for FRB 20190110C). For cases (ii) and (iii), we quote values for at least 3 and 8 galaxies, respectively, because that was the total number of galaxies with those brightnesses found in the combined localization regions of all 13 FRBs.  These numbers indicate that although there is a relatively low $P_\mathrm{cc,gal}$ of finding a bright galaxy in the localization region of an individual FRB (e.g. $1.0\%$ for FRB 20200223B). When considering the search region as a whole, we do not see evidence for a strong overabundance of bright galaxies within the \emph{combined} 13 CHIME/FRB repeaters regions. This implies that we cannot make associations for our galaxies using $P_\mathrm{cc}$ based on galaxy number counts \emph{alone}.

\subsection{Summary of Possible Host Associations} \label{subsec:summary-robust-host} 

In the analysis above, we identified three galaxies that we felt were worthy of further investigation due to \texttt{PATH} strongly preferring an individual galaxy over any other source \emph{detected} in the field. We found that when ``look-elsewhere'' effects were considered, there is not a \emph{strong} overabundance of bright galaxies ($m_r < 18$ mag) within the 13 combined CHIME/FRBs localization regions. However, if we assume that FRBs must come from galaxies, then the Bayesian approach of \texttt{PATH}---which takes into account not only galaxy number counts but also information on transients offsets---offers a means to robustly consider the relative probabilities of possible hosts for an individual FRB. We, therefore, rely primarily on the \texttt{PATH} analysis. Here, we synthesize information available for the three galaxies necessary to infer whether a likely host association can be made.

\begin{figure*}
    \centering
    \includegraphics[width=0.90\textwidth,height=0.2\textheight]{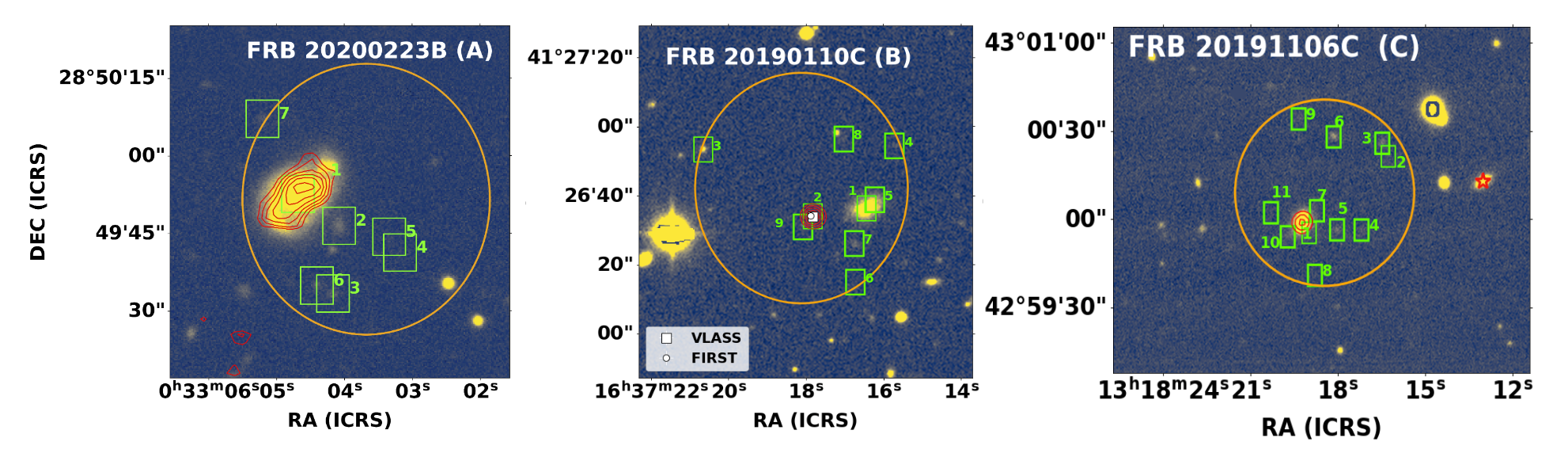}
    
    \caption{DESI $r$-band images showing the 90\%  baseband localization region of FRBs 20200223B, 20190110C and 20191106C (orange ellipses). Green boxes show the positions of host galaxy candidates. The red background contours are from high-resolution LOFAR Two-metre Sky Survey \cite[LoTSS;][]{Shimwell2022} radio data. \emph{Left:} FRB 20200223B: There are 7 sources within the localization region from the DESI catalog. Galaxy A, marked as 1 in the image, is a spiral galaxy at $z_\mathrm{spec} = 0.06$ and is the most plausible host galaxy of FRB 20200223B. The red background contours from LoTSS fit the morphology of A and are thus consistent with star formation activities of the galaxy.  \emph{Middle:} FRB 20190110C: We present 9 sources from the DESI catalog in this FRB region. Galaxy B, marked as 1 is the plausible host galaxy of FRB 20190110C at $z_\mathrm{spec} = 0.122$. The white markers represent VLASS and FIRST radio point sources in the field of the FRB spatially coincident with another nearby galaxy named source 2. There is no radio source that is coincident with galaxy B. \emph{Right:} FRB 20191106C: There are 11 sources within the localization region from the DESI catalog. Source 1 (C) is a galaxy at $z_\mathrm{spec} = 0.108$ and is the most likely host of FRB 20191106C. There is a LoTSS radio source just outside the region shown as a red star but the one that is spatially coincident with galaxy C is shown as a red contour.}
    \label{fig:R49sdss-C64sdss-image}
\end{figure*}

\emph{FRB 20200223B: } There are seven extended sources in the DESI catalog in the field of FRB 20200223B. These are marked as green boxes in the left panel of Figure~\ref{fig:R49sdss-C64sdss-image}, and have apparent $r$-band magnitudes in the range 16.08 $<$ m$_{\mathrm{r}}$ $<$ 23.97. The morphology of Source 1 suggests a spiral galaxy while sources 2-7 are faint objects in the field. Of these, Source 1 (SDSS J003304.68$+$284952.6; hereafter A) with m$_{\mathrm{r}}$(AB) $=$ 16.080$\pm$0.001\,mag and an effective radius of 5.81$\pm$0.02\,arcseconds is the galaxy preferred by \texttt{PATH} (see Table \ref{tab: path_A} in Appendix~\ref{appendix} for \texttt{PATH} probabilities assigned to all galaxies in the field of this FRB).

When adopting $\mathrm{P(U)} = 0$, \texttt{\texttt{PATH}} gives a probability of 99.4\% that the FRB is associated with galaxy A (Table~\ref{tab:catalogue details}). Given that the DESI depth corresponds to an absolute magnitude of roughly $-$15.90\,mag at $z_\mathrm{max} < $0.19, our initial search should have been sensitive to all FRB hosts discovered to date. The probabilities obtained using the prior, $\mathrm{P(U)} = 0$ are likely reasonable for this FRB.
In addition, we found that the probability of finding an m$_{\mathrm{r}} = $16.1\,mag galaxy in the CHIME localization region for FRB 20200223B by chance was $\sim$1.0\%. 

Finally, we note that the photometric redshift for this galaxy ($z_\mathrm{phot} = $0.06$\pm$0.01) is less than $z_\mathrm{max}$ for this FRB ($z_\mathrm{max}$ $<$ 0.19). We, therefore, consider this host association robust. A is located at $\alpha$(J2000) $=$ 00$^\mathrm{h}$33$^\mathrm{m}$04\fs68 (0.015$\arcsec$) and $\delta$(J2000) = $+$28\degr49\arcmin52\farcs60 (0.021$\arcsec$) \citep{sdss2015}. 

\emph{FRB 20190110C: }  We found nine objects that are not point sources in the field of FRB 20190110C by searching the DESI DR9 catalog (shown in green boxes on the middle panel of Fig ~\ref{fig:R49sdss-C64sdss-image}). The nine sources have apparent DESI $r$-band magnitudes in the range 18.0 $<$ m$_{\mathrm{r}}$ $<$ 23.9. By visual inspection, source 1 looks like an irregularly shaped object with an extended clump classified as source 5 by DESI. We, therefore, consider them the same as opposed to distinct sources. Source 1, SDSS J163716.43$+$412636.2 (hereafter, B) has a probability of being associated with the FRB of $P_{\mathrm{\texttt{PATH}}}$ $=$ 91.8\% (when adopting $\mathrm{P(U)} = 0$) and has a $z_\mathrm{phot} = $0.10 $\pm $0.02 compared to $z_\mathrm{max}$ $<$ 0.22 for this target. It has apparent r-band magnitude, m$_{\mathrm{r}}$(AB) 18.009 $\pm$ 0.006 and an effective radius of 3.38 $\pm$ 0.03\,arcseconds. 

The DESI depth of m$_{\mathrm{r}}$(AB) $\sim$24\,mag corresponds to an absolute magnitude of roughly $-$16.26\,mag at $z_\mathrm{max} < $0.22, and thus we were sensitive to all FRB hosts discovered to date. Given this, the probabilities obtained using the prior, $\mathrm{P(U)} = 0$ is likely reasonable for this FRB. In addition, we found that the probability of finding an m$_{\mathrm{r}} = 18.0$\,mag galaxy in the CHIME localization region by chance was 10.2\%. We, therefore, conclude that B is the likely host galaxy of FRB 20190110C. Galaxy B is located at $\alpha$(J2000) $=$ 16$^\mathrm{h}$37$^\mathrm{m}$16\fs43 (0.022$\arcsec$) and $\delta$(J2000) = $+$41\degr26\arcmin36\farcs30 (0.024$\arcsec$) \citep{sdss2015}. We report the \texttt{\texttt{PATH}} probability of all the sources in the field of this FRB in Table \ref{tab: path_B} of Appendix~\ref{appendix}.

\emph{FRB 20191106C: } 11 objects that are not point sources were found in the DESI catalog within the baseband region of FRB 20191106C (See the green boxes in Figure~\ref{fig:R49sdss-C64sdss-image}.  The eleven sources have apparent DESI $r$-band magnitudes in the range 17.3 $<$ m$_{\mathrm{r}}$ $<$ 23.9.  

We note that the DM for FRB 20191106C is larger than those of the other two FRBs discussed in this section, leading to a higher $z_{\mathrm{max}}$. As a result, the DESI limit corresponds to a slightly higher absolute magnitude of roughly $-$17.48\,mag at $z_\mathrm{max} < 0.36$ and it is possible that some dwarf galaxies were missing from our initial candidate list (for example, the known host of FRB 121102 at $-16.5$ mag; \citealt{Chatterjee_2017}). Given these, it is likely more appropriate to use $\mathrm{P(U)} = 0.1$ for this FRB. 

Even when adopting $\mathrm{P(U)} = 0.1$, \texttt{PATH} significantly prefers source 1 (SDSS J131819.23+425958.9, hereafter, C) over any other objects in the field. It assigns a probability of 81.5\% to galaxy C compared to $\lesssim$1\% for any other galaxy in the field (see Table~\ref{tab: path_C} in Appendix~\ref{appendix} where we report the \texttt{PATH} probability of all the sources in the field of this FRB).  However, we found that the probability of finding an m$_{\mathrm{r}} = $17.3\,mag galaxy in the CHIME localization region by chance is 6.3\%. Therefore, we argue that C is likely a good candidate host of the FRB. It is located at $\alpha$(J2000) $=$ 13$^\mathrm{h}$18$^\mathrm{m}$19\fs23 (0.01$\arcsec$) and $\delta$(J2000) = $+$42\degr59\arcmin58\farcs97 (0.01$\arcsec$) \citep{sdss2015} and looks like an elliptical/irregular galaxy with m$_{\mathrm{r}}$(AB) of 17.306 $\pm$ 0.003.

\emph{Summary:} Based on the results above, we conclude that Galaxies A and B are highly probable hosts for FRB 20200223B and FRB 20190110C, respectively. Both galaxies have $>$90\% probability of being the true host based on \texttt{PATH} and the depth of the archival surveys is sufficient to detect all previously discovered FRB host galaxies. As described above, although it is not possible to make host associations based on galaxy number counts in the 13 CHIME/FRB localization regions alone, if we assume that FRBs come from galaxies, the Bayesian approach of \texttt{PATH} provides a means to assess the most likely host. In contrast, the association of galaxy C to FRB 20191106C is more marginal due to the shallower depth of archival surveys which means that the probability of unseen galaxies as the host is non-negligible. This leads to a \texttt{PATH} probability below 90\% when adopting a value of P(U) that is greater than 0. However, given that \texttt{PATH} strongly preferred this host over any other source found in the field, we chose to also describe the properties of this galaxy in the section below.

Finally, we note that the FRB host galaxy association method described in this manuscript is different from that used in previously published CHIME/FRB papers \citep{Bhardwaj_2021, BhardwajR42021, Michilli2022}. The latter involves obtaining spectroscopic redshifts of all the galaxy candidates within the FRB field and uses the $z_\mathrm{spec}$ $< z_\mathrm{max}$ condition to decide on host association. Given that we are looking at many FRB fields, some of which have high DMs, it is beyond the scope of the present work to conduct spectroscopic analysis for all the candidates in each field. Hence, we adopt the probabilistic method as described here. However, future more detailed studies involving spectroscopic redshifts could be of use.

\section{Observations of Likely Host Associations} \label{sec:observations-likely-host}
Here we describe the full set of archival and follow-up observations obtained for the plausible host galaxies of FRBs 20200223B, 20191106C, and 20190110C. These will be used to measure detailed host properties in \S \ref{sec:properties-of-host}.

\subsection{Optical and Infrared Photometry} \label{subsec:Optical and IR Photometry}
We retrieved photometric data for each of the 3 host galaxies from SDSS, DESI, the Wide-field Infrared Survey Explorer (WISE) \citep{Wright2010}, 2MASS \citep{Two2003}, the Galaxy Evolution Explorer (GALEX) \citep{Bianchi2017} and the Spitzer Enhanced Imaging Product (SEIP) source list (SSTSL2) \citep{Spitzer2021}. We will use these optical-infrared spectral energy distributions to estimate various properties of the galaxies in \S \ref{subsec:prospector}. We corrected all data for Galactic foreground extinction using the dust reddening map of \cite{Schlegel1998A} and assuming an R$_{\mathrm{v}} = $3.1 Milky Way extinction curve. $E(\bv)$ values are 0.0489(8), 0.0081(5), and 0.0200(8) for galaxies A, B, and C respectively.

\begin{figure}
    \centering
    \includegraphics[width=0.50\textwidth]{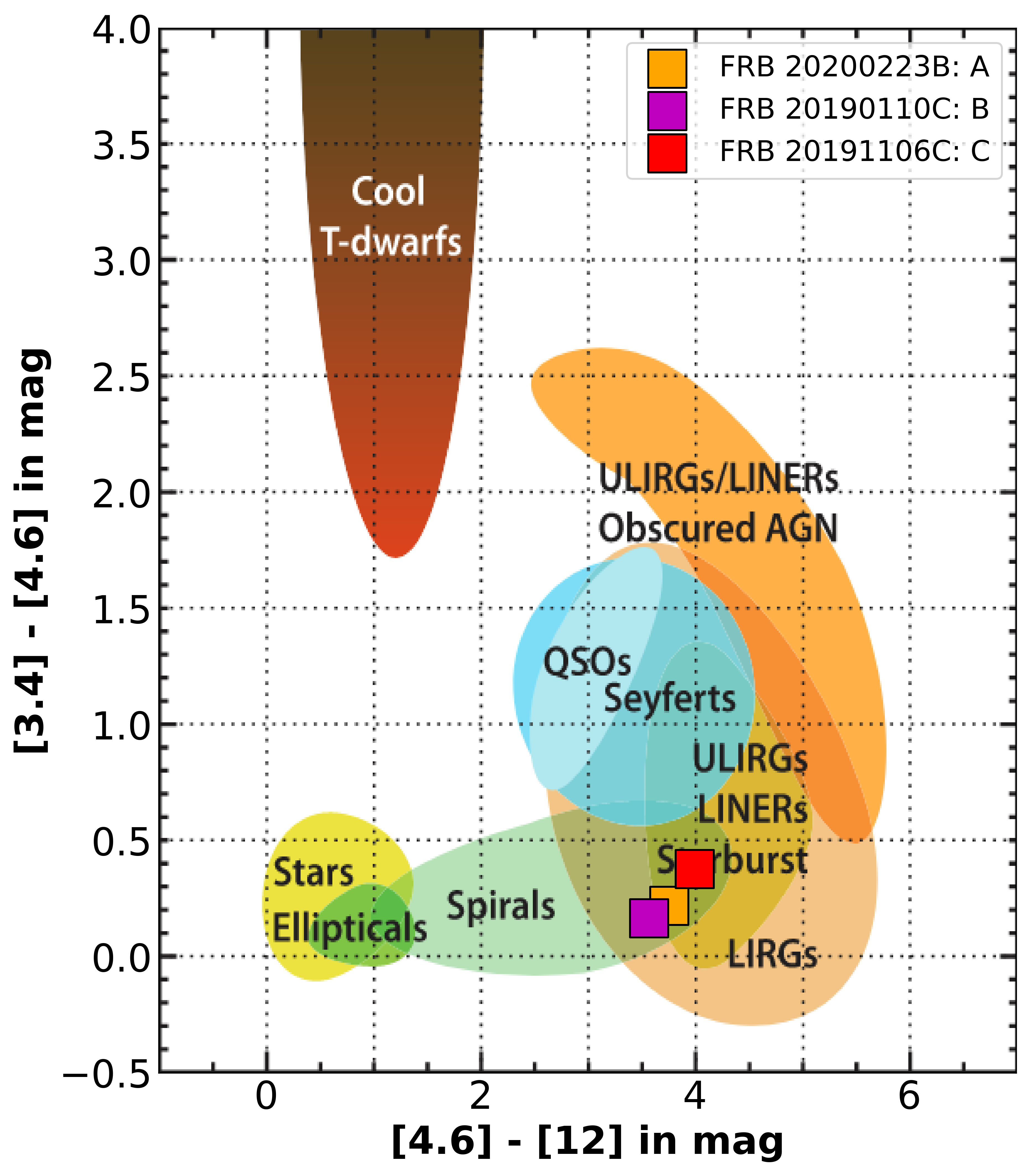}
    \caption{WISE diagnostic plot adapted from \cite{Wright2010} showing the location of various classes of objects with the 3 plausible host galaxies over-plotted as orange, purple, and red squares for galaxies A, B, and C. The infrared colors from the 3 plausible host galaxies all seem to be associated with various types of star-forming galaxies.}
    \label{fig:wise-plot}
\end{figure}

The infrared colors of galaxies A, B, and C in the WISE bands can be seen in Figure \ref{fig:wise-plot}.  Using the WISE color-map diagnostic plot \citep{Wright2010,nik14}, galaxy A falls in a region also occupied by spiral, star-forming, starburst, low-ionization nuclear emission-line regions (LINER) galaxies, and luminous infrared galaxies (LIRG) \citep{Jarett2017WISE}. While the morphology of galaxy B is ambiguous, the infrared colors also overlap with spiral galaxies and LIRGs, implying a star-forming galaxy.
Finally, we note that Galaxy C was previously classified as an elliptical galaxy by \cite{Kuminski2016} with $\sim$88\% certainty and an ellipticity of 0.4 \citep{Simard2011}. However, the infrared colors overlap with various classes of star-forming galaxies (spiral, starburst, and LIRGs), which is inconsistent with a typical quiescent elliptical galaxy. Moreover, this infrared color result is in line with the result of \cite{Ellison2016} who use artificial neural networks to model the total infrared luminosity, where they predict that C is a star-forming galaxy. We conclude that C could be among the rare population of elliptical galaxies with ongoing star formation.

\subsection{Optical Spectroscopy} \label{subsec:Optical-Spectroscopy}
For all three likely host galaxies, we use optical spectroscopy to determine their spectroscopic redshifts and emission line fluxes--- particularly, H$_{\mathrm{\alpha}}$, H$_{\mathrm{\beta}}$, [OIII]($\lambda 5006\AA$), and [NII]($\lambda 6582\AA$)--- that will then be used to estimate other galaxy properties in \S \ref{subsec:emission-line}. In all cases, we convert redshifts to luminosity distances assuming a flat $\Lambda_{\mathrm{\textit{CDM}}}$ cosmology with $H_{\mathrm{0}}$ $=$ 67.7 km\,s$^{-1}$ Mpc$^{-1}$, $\Omega_{\mathrm{m}}$ = 0.31, and $\Omega _{\mathrm{\wedge}}$ = 0.70 \citep{planck2018vi}.

\subsubsection{Archival Spectroscopy} \label{ssubsec: archival-spec}
For galaxies A and C, archival spectra are available. For FRB 20200223B, the likely host galaxy, A, was observed and released as part of the Large Sky Area Multi-Object Fiber Spectroscopic Telescope \citep[LAMOST;][]{cui2012} Data Release 5 (DR5). We retrieve the redshift and relevant emission line fluxes. The spectroscopic redshift is $z_\mathrm{spec} =$ 0.06024(2) which corresponds to a luminosity distance of 278.97(9)\rm{Mpc}. For FRB 20191106C, the likely host galaxy, C, was observed as part of SDSS DR12 and the redshift was reported by \cite{sdss2015}. The spectroscopic redshift is $z_\mathrm{spec} =$ 0.10775(1), which corresponds to a luminosity distance of 515.32(4) \rm{Mpc}. Emission line fluxes for galaxy C were extracted from the SDSS-III spectroscopic survey \citep{Eisenstein2011,Ahumada2020}. 

For both A and C, we correct the previously published LAMOST/SDSS line fluxes for foreground Milky Way extinction. We determine correction factors based on the \cite{Cardelli1989} extinction law (R$_{\mathrm{V}}$ $=$ 3.1) at the rest wavelengths of H$_{\mathrm{\alpha}}$, H$_{\mathrm{\beta}}$, [OIII]($\lambda 5006\AA$) and [NII]($\lambda 6582\AA$). In addition, we also check for intrinsic host galaxy extinction by calculating the Balmer decrement from the H$_{\mathrm{\alpha}}$, H$_{\mathrm{\beta}}$ line fluxes. Compared to the theoretical Case-B recombination line ratio of H$_{\mathrm{\alpha}}$/H$_{\mathrm{\beta}}$ $=$ 2.86, we do not find evidence for significant additional extinction.

\subsubsection{Gemini North Spectroscopy} \label{ssubsec: gemini-spec}

For FRB 20190110C, there is no archival spectroscopy available for its plausible host galaxy, B. Hence we obtained an observation with the Gemini Multi-Object Spectrograph (GMOS) of the Gemini-North telescope in Hawaii. The data were obtained on 2022 August 20 during semester 2022B. We used the 1.0" long-slit, R400 grating and OG515 ($>$ 520\,nm) blocking filter, at central wavelengths of 720/730\,nm. In total, we obtained five 1200\,second exposures with a mean airmass of 1.2.

The data were reduced using the standard packages of \texttt{gmos} and \texttt{gemini} routine in \texttt{PYRAF} \citep{pyraf2012}. This included overscan correction, flat fielding, sky subtraction, wavelength calibration, extraction, and flux calibration. Chip gaps and cosmic rays were removed from individual exposures and the different exposure spectra were combined to produce a single spectrum. In order to estimate the redshift of galaxy A from the spectrum, we measured the observed central wavelengths of the H$_{\mathrm{\alpha}}$, H$_{\mathrm{\beta}}$, [NII], and SII emission lines. We recorded a redshift of $z_\mathrm{spec} =$ 0.12244(6), which corresponds to a luminosity distance of 591.2(3) \rm{Mpc}. 
 
 The spectrum was then corrected for cosmological expansion using the redshift and corrected for Milky Way extinction using the \cite{Cardelli1989} extinction law. As above, we find no evidence for significant additional internal extinction when calculating the Balmer decrement. The line fluxes of H$_{\mathrm{\alpha}}$, H$_{\mathrm{\beta}}$, [OIII]($\lambda 5006\AA$), and [NII]($\lambda 6582\AA$) were then measured by fitting a gaussian profile to the rest-frame spectrum. The process was carried out in an automated manner and repeated 100 times to estimate the error in the line fluxes due to uncertainty at the continuum level.

  \begin{figure}
    \centering
    \includegraphics[width=0.50\textwidth]{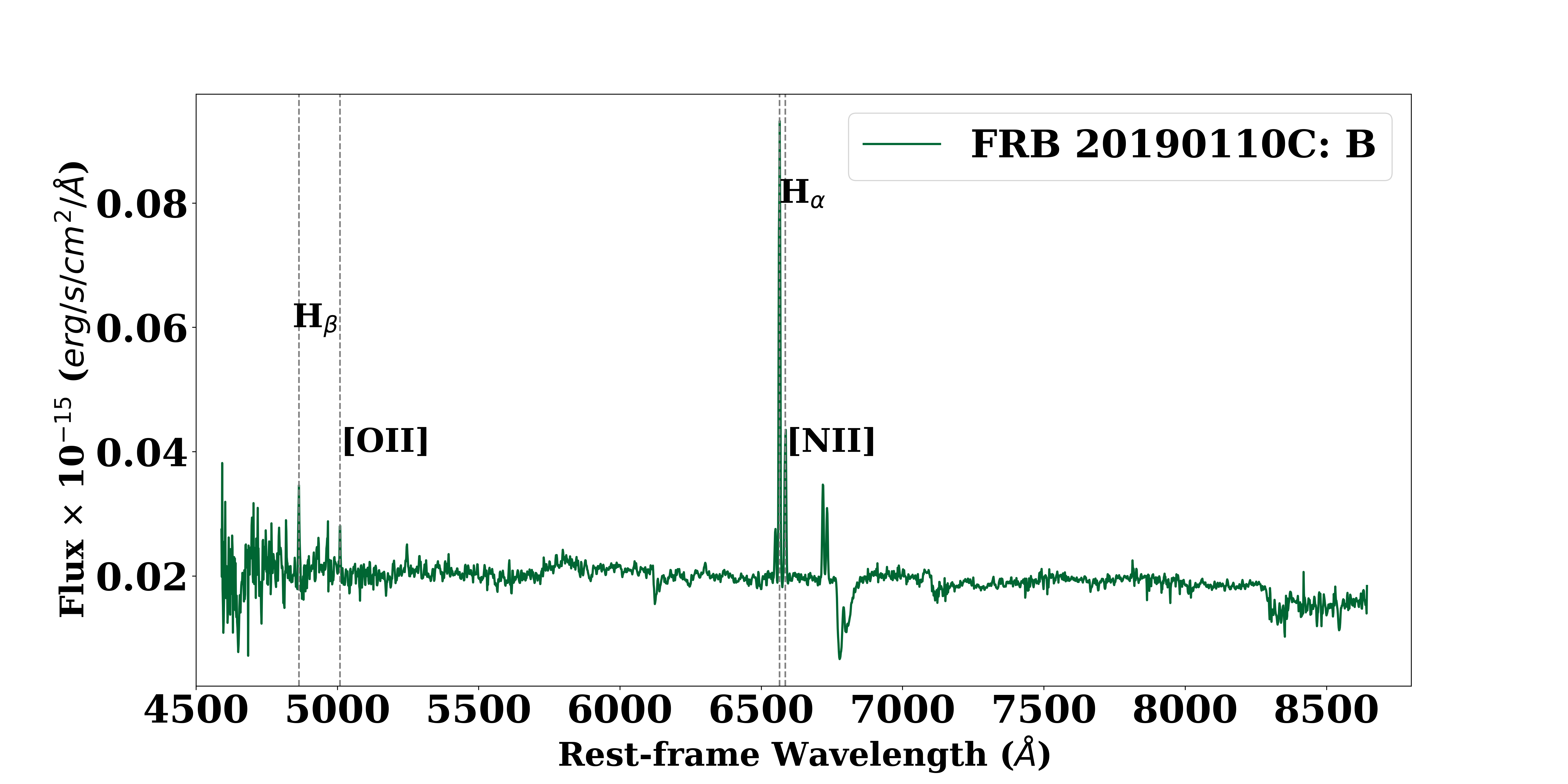}
    \caption{Gemini North spectra for the plausible host galaxy of FRB 20190110C, B. Gray lines highlight the location of prominent emission lines that were used for analysis.}
    \label{fig:c64-spectra}
\end{figure}

\subsection{Radio Emission} \label{subsec:Radio-Emission}
We searched for the presence of a persistent radio source (PRS) within the error regions of the three FRBs using archival radio catalogs from the VLA Sky Survey \citep[VLASS;][]{Lacy_2020}, the NRAO VLA Sky Survey \citep[NVSS;][]{Condon1998}, the Faint Images of the Radio Sky at Twenty-Centimeters \cite[FIRST;][]{Becker1995} survey, the TIFR-GMRT Sky Survey \cite[TGSS;][]{Intema2017} and the high-resolution component of the LOFAR Two-metre Sky Survey \cite[LoTSS;][]{Shimwell2022}. 

There are LoTSS radio sources at the locations of the host galaxies of FRB 20200223B and FRB 20191106C, with integrated flux densities of 2.96 $\pm$ 0.59\,mJy and 3.29 $\pm$ 0.14\,mJy, respectively, at a frequency of 144\,MHz. We measured upper limits on the presence of a radio source at the location of B, which is the plausible host of FRB 20190110C. In Table ~\ref{tab:radio-catalogue details}, we report detections and upper limits on radio emission in the vicinity of the plausible host galaxies from these datasets. In \S \ref{subsec:prs-constraint}, we will investigate the possible origin of any emission detected.

\begin{deluxetable*}{ccccccccc}
\tabletypesize{\small}
\tablecaption{Summary of PRS searches for the host galaxies of FRB 20200223B (A), FRB 20190110C (B) and FRB 20191106C (C) }\label{tab:radio-catalogue details}

\tablehead{\colhead{} & \colhead{} & \colhead{} & \colhead{A} & \colhead{B} & \colhead{C} & \colhead{}\\
\colhead{Survey} & \colhead{Frequency} & \colhead{Angular} & \colhead{Flux density} & \colhead{Flux density} & \colhead{Flux density} & \colhead{References} \\
\colhead{} & \colhead{(\,MHz)} & \colhead{Resolution (\,$^{\prime\prime}$)} & \colhead{(\,mJy)} & \colhead{\,mJy} & \colhead{\,mJy}  & \colhead{}
  }
\startdata
\centering
      FIRST &  1400 & 5 & $<$0.75$^a$ & $<$0.75$^a$ & $<$0.75$^a$ & \cite{Becker1995} \\
      NVSS &  1400 & 45 & $<$12.5$^a$ & $<$12.5$^a$ & $<$12.5$^a$ & \cite{Condon1998} \\
      VLASS &  3000 & 2.5 & $<$0.69$^a$ & $<$0.60$^a$ & $<$0.63$^a$ & \cite{Lacy_2020} \\
      TGSS ADR1 &  150 & 25 & $<$17.5$^a$ & $<$17.5$^a$ & $<$17.5$^a$ & \cite{Intema2017} \\
      LoTSS &  144 & 6 & 2.96 $\pm$ 0.59 & $<$0.42$^a$ & 3.29 $\pm$ 0.14 & \cite{Shimwell2022} \\   
\enddata
\textbf{Notes.} \\
    $^a$ 5$\sigma$ RMS sensitivity or upper limit.
\end{deluxetable*}
 
\section{Properties of Likely Host Associations} \label{sec:properties-of-host}

Here we described the properties of galaxies A, B, and C. We investigate the implication of the results from optical spectroscopy used to estimate galaxy properties of A, B, and C in \S \ref{subsec:emission-line}. We then report on the SED fitting using \texttt{\texttt{Prospector}} for galaxies A and B in \S \ref{subsec:prospector}, and place constraints on the presence of radio emission from these galaxies in \S \ref{subsec:prs-constraint}. 
We check for any intervening structures that could contribute to the DM and discuss their effect on DM$_{\mathrm{host}}$ for the FRB in our sample in \S \ref{subsec:implication-DM-host}. Finally, in \S~\ref{subsec:comparison-to-other-FRBs} we compare the properties of these galaxies to the hosts of other FRBs.

\begin{table*}[t]
\begin{center}
\caption{Properties of the plausible host galaxies of FRB 20200223B (A), FRB 20190110C (B), and FRB 20191106C (C)}
\begin{tabular}{lccc} \hline
    Observed property & A Values & B Values & C Values \\ \hline
    Right Ascension (J2000) & 00$^\mathrm{h}$33$^\mathrm{m}$04\fs68 (0.015$\arcsec$) & 16$^\mathrm{h}$37$^\mathrm{m}$16\fs43 (0.022$\arcsec$) & 13$^\mathrm{h}$18$^\mathrm{m}$19\fs23 (0.01$\arcsec$) \\
    Declination (J2000) & $+$28\degr49\arcmin52\farcs60 (0.021$\arcsec$) & $+$41\degr26\arcmin36\farcs30 (0.024$\arcsec$) & $+$42\degr59\arcmin58\farcs97 (0.01$\arcsec$) \\
    Galaxy name & SDSS J003304.68$+$284952.6 & SDSS 163716.43$+$412636.2 & SDSS J131819.23+425958.9 \\
    Galactic Longitude ($l$) and Latitude ($b$) & 118.08, $-$33.86 & 65.57, $+$42.09 & 105.68, $+$73.22 \\
    Apparent r-band mag (AB) & $16.080\pm0.001$  & 18.009$\pm$0.006 & 17.306$\pm$0.003 \\
    $E(\bv)$ (mag) & 0.0483(8)  & 0.0078(5)  & 0.0197(8) \\

    spectroscopic redshift, $z_\mathrm{spec}$ & 0.06024(2)  & 0.12244(6) & 0.10775(1) \\
    Luminosity distance (Mpc) & 278.97(9)  & 591.2(3) & 515.32(4)\\ 
    Effective radius, R$_\mathrm{eff}$ (kpc)$^{e}$ & 7.83$\pm$0.03 & 9.65$\pm$0.09 & 4.01$\pm$0.01\\
    M$_{\mathrm{r}}$ & $-$21.15$\pm$0.01 & $-$20.71$\pm$0.02 & $-$21.20$\pm$0.01 \\
    M$_{\mathrm{u}}$ $-$ M$_{\mathrm{r}}$ & 1.84$\pm$0.05 & 1.64$\pm$0.11 & 2.08$\pm$0.07 \\
    
    H$_{\mathrm{\alpha}}$ & 8.63$\times$10$^{-15}$$^{f}$ & 4.8042(8)$\times$10$^{-16}$ & 6.09$\times$10$^{-15}$$^{f}$  \\
    
    H$_{\mathrm{\beta}}$ & 1.17$\times$10$^{-15}$$^{f}$ & 6.43(8)$\times$10$^{-17}$ & 8.88$\times$10$^{-16}$$^{f}$ \\
    
    [OIII]($\lambda 5006\AA$) & 2.94$\times$10$^{-15}$$^{f}$ & 2.93(7)$\times$10$^{-17}$ & 1.27$\times$10$^{-16}$$^{f}$  \\
    
    [NII]($\lambda 6582\AA$) & 4.17$\times$10$^{-15}$$^{f}$ & 1.703(9)$\times$10$^{-16}$ & 2.09$\times$10$^{-15}$$^{f}$ \\

    \hline Inferred Property &  &  & \\ \hline
    
   SFR$_{\mathrm{H_{\mathrm{\alpha}}}}$ ($M_{\mathrm{\odot}}$ yr$^{-1}$)$^{a}$ & N/A &0.1575(6) & 1.53$^{f}$\\
  Oxygen Abundance [$Log_{\mathrm{10}}$ (O/H)]$^{a}$ + 12 & N/A & 9.102(2) & 9.10$^{f}$ \\

     SFR ($M_{\mathrm{\odot}}$ yr$^{-1}$)$^{b}$ & 0.59$_{-0.04}^{+0.04}$ &  0.54$_{-0.04}^{+0.04}$ & 4.75$_{-1.27}^{+1.29}$$^c$ \\

    Stellar mass (M$_{\mathrm{\odot}}$)$^{b}$ & $5.6_{-0.93}^{+1.14}$ $\times$ $10^{10}$ & $2.5_{-0.17}^{+0.10}$ $\times$ $10^{10}$ & $4.5_{-1.2}^{+1.2}$ $\times$ $10^{10c}$\\
    Mass weighted Age (Gyr)$^{b}$ & 3.37$_{-0.73}^{+0.47}$ & 2.59$_{-1.24}^{+0.80}$ & - \\
    Specific star formation rate [log(sSFR)] (yr$^{-1}$) & $-$10.98$_{-0.03}^{+0.00}$$^{d}$  & $-$10.66$_{-0.14}^{+0.07}$$^{d}$ & $-$10.61$_{-0.21}^{+0.14}$$^{c}$ \\ 
   
 \hline
 
\end{tabular}
\label{tab: A-galaxy-properties}
\end{center}
 \textbf{Notes.} \\
    $^a$Values obtained using the optical spectrum of the host galaxies. \\
    $^b$Value estimated from \texttt{Prospector}.\\
    $^c$ Published values obtained from literature \citep{Chang2015}. \\
    $^d$ Values obtained using SFR and the stellar mass values from the \texttt{Prospector} analysis.\\
    $^{e}$ Values are half-light radius taken from DESI legacy catalog \citep{dsl+19} \\
   $^{f}$ Values published without errors. Our analysis of these data shows that the errors will not change the result of our analysis.\\
    N/A represents parameter estimates that are not reliable, and hence not published.
\end{table*}

\subsection{Optical Emission Line Diagnostics} \label{subsec:emission-line}

For the three host galaxies, we use the corrected emission line fluxes (see \S \ref{subsec:Optical-Spectroscopy}) to infer their dominant source of photo-ionization, star formation rates, and metallicities.

\begin{figure}
    \centering
    \includegraphics[width=0.50\textwidth]{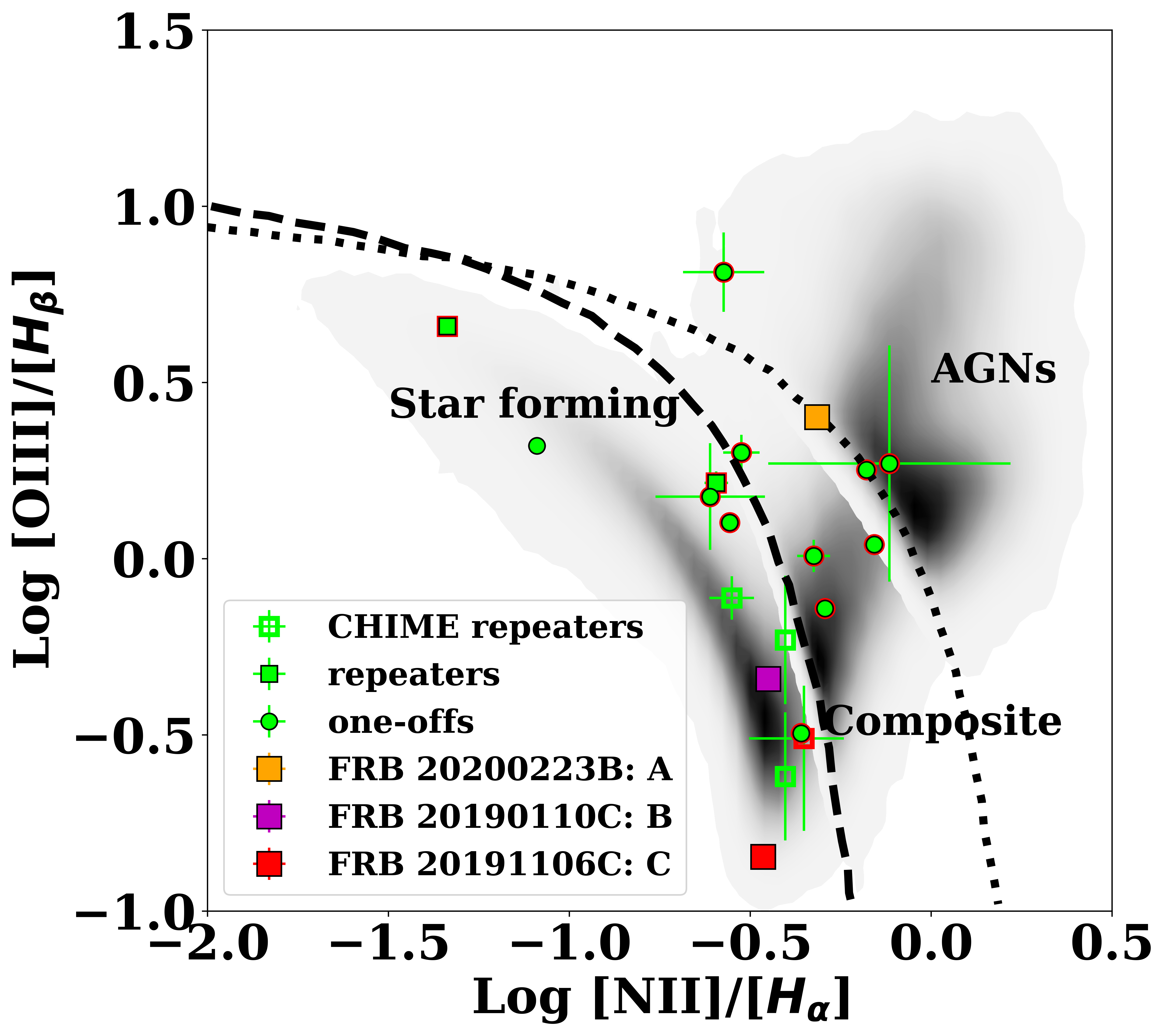}
    \caption{BPT diagram showing the dominant source of ionization using the nebular emission line ratios from the 3 plausible host galaxies and other previously published FRB hosts. The greyscale background represents the density of galaxies from SDSS DR8 \citep{Aihara2011}. The dashed line \citep{Kauffmann2003} indicates the empirical division between star-forming galaxies and AGNs while the dotted line \citep{Kewley2001} shows the theoretical demarcation. FRB repeaters and one-offs are represented with squares and circles respectively while the ones with robust host associations ($\sim <$1 arcsecond) have red edge color. The hollow green squares are other CHIME/FRB repeaters while the filled red, purple, and orange squares indicate the 3 host galaxies proposed here for CHIME/FRB repeaters. The location of the two galaxies (B and C) clearly suggests star formation while the location of A indicates an AGN contribution to the line fluxes. The representative repeating and non-repeating FRBs' line ratios taken from \cite{Heintz2020, Bhandari2021, Michilli2022, BhardwajR42021, Fong2021, Mannings2021}, indicates that most FRB host galaxies, especially repeaters, are mostly star-forming galaxies. See \S \ref{subsec:comparison-to-other-FRBs} for details.}
    \label{fig:BPT-plot}
\end{figure}

\emph{BPT diagram:} In order to probe the major source of photo-ionization in each of the three galaxies, we first placed all 3 of them on the updated ``Baldwin, Phillips \& Terlevich'' \citet{BPT1-1981} (BPT) diagram of \cite{BPT22011} shown in Figure \ref{fig:BPT-plot}. Galaxies A, B, and C are designated by orange, magenta, and red squares, while in the background we plot the density distribution of nearby (0.02 $< z <$ 0.35) emission line galaxy ratios with S/N $>$5 from the SDSS spectroscopic data from Data Release 8 \citep[DR8:][]{Aihara2011}. Comparison with other overplotted FRB hosts line ratios is made in \S \ref{subsec:comparison-to-other-FRBs}.

From this, we see that B and C galaxies fall within the region where the optical emission line ratios are dominated by star formation, while A appears to have a non-negligible contribution from an AGN to its optical line ratios. We note that there are no error bars associated with emission line fluxes for galaxies A and C (which were taken from LAMOST and SDSS, respectively, see above). From examining these archival spectra, we find that both $H_{\mathrm{\alpha}}$ and [NII] are strongly detected while [OIII] (and $H_{\mathrm{\beta}}$ in the case of galaxy A) fall in noisy regions of the spectra. However, if we assume a  conservative error of 30\% (corresponding to a 3$\sigma$ detection of these lines), the resulting errors would not be significant enough to shift either galaxy from the star-forming or composite region of the BPT diagram, respectively. As a result, we consider these broad classifications robust and calculate other properties of star formation in the host galaxies (star formation rate, metallicity) for B and C only. We do not calculate the star formation rate and metallicity for A due to the AGN contamination.

\emph{Star formation rates:} In the absence of internal extinction as was evident from the Balmer decrement of each of the three galaxies, we calculate star formation rates from the H$_{\mathrm{\alpha}}$ line fluxes for B and C without Balmer decrement correction using the equations from \cite{kennicutt1998}: SFR($H_{\mathrm{\alpha}}$) $=$ 7.9$\times$10$^{-42}$ (L$_{\mathrm{H_{\mathrm{\alpha}}}}$)
From this we find star formation rates of 0.1575(6)\,$M_{\mathrm{\odot}}$ yr$^{-1}$ at L$_{\mathrm{H_{\mathrm{\alpha}}}}$ = 1.994(7) $\times$ 10$^{40}$\,ergs\,s$^{-1}$ for B and 1.53\,$M_{\mathrm{\odot}}$ yr$^{-1}$ at L$_{\mathrm{H_{\mathrm{\alpha}}}}$ = 1.94 $\times$ 10$^{41}$\,ergs\,s$^{-1}$ for C (See Table \ref{tab: A-galaxy-properties}). These values are consistent with galaxies with active star formation.

\emph{Metallicity:} We calculate metallicities using Equation 1 from \cite{Pettini2004} (12 $+$ log (O/H) $=$ 8.90 $+$ 0.57 $\times$ N2), where N2 $=$ [NII]$\lambda$6583/H$_{\mathrm{\alpha}}$. Using these ratios to estimate the metallicity of the galaxy resulted in an oxygen abundance of 9.102(2) which is equivalent to 2.76(1) Z$_{\mathrm{\sun}}$ metallicity for B, and 9.10 which is equivalent to 2.7 Z$_{\mathrm{\sun}}$ metallicity for C. These are high metallicities expected from an old population.

\subsection{\texttt{Prospector} SED Fitting} \label{subsec:prospector}
We fitted the observed photometric data and spectroscopic redshifts using the stellar population synthesis modeling code \texttt{Prospector} \citep{prospector2021}, in order to determine the stellar mass of galaxies A and B (the mass of Galaxy C properties was previously published by \citealt{Chang2015}, hence there was no need to carry out SED fitting for it). \texttt{Prospector} uses the stellar population synthesis library provided in the Flexible Stellar Populations Synthesis (FSPS) stellar population code \texttt{python-fsps} \citep{Conroy_2009} to fit the observed data. We assumed a Chabrier initial mass function \citeyearpar{Chabrier2003} and allowed a delayed-$\tau$ star formation history (SFH) model \citep{Carnall_2019} that has 5 free parameters (stellar mass, stellar metallicity, galaxy age, dust attenuation optical depth for a foreground screen, and star formation timescale for an exponentially declining SFH). Reasonable priors as listed in Table \ref{tab:prospector-prior} were set for the free parameters. The likelihood given the free parameters was then sampled using a Markov Chain Monte Carlo (MCMC) approach through the \texttt{emcee} code \citep{Foreman-Mackey2013} and the posterior distributions were calculated. The extinction-corrected photometry was converted to flux densities as listed in Table \ref{tab:prospector-sed}. 

\begin{figure*}
    \centering
    \includegraphics[width=0.45\textwidth]{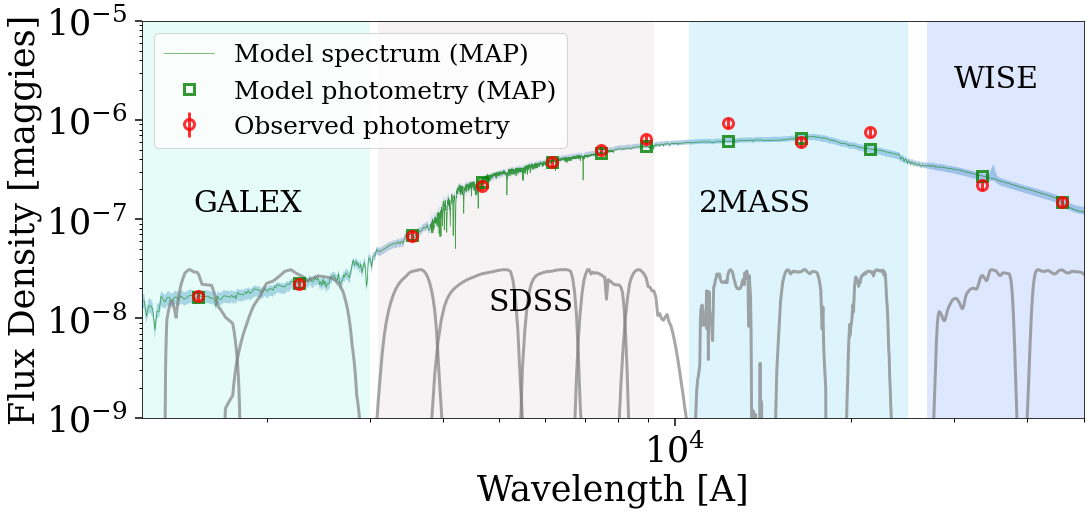}
        \includegraphics[width=0.45\textwidth]{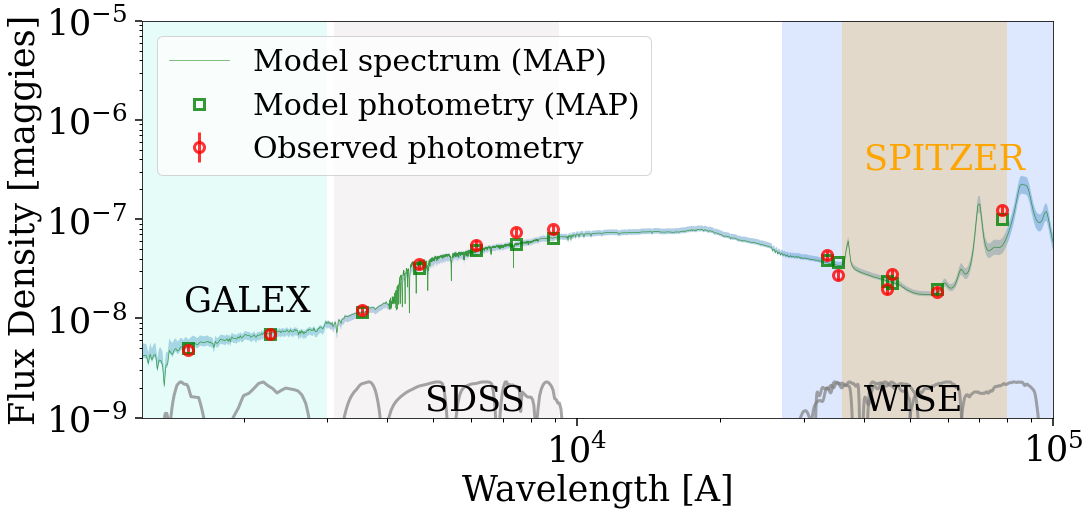}
    \caption{\emph{Left:} Best-fit \texttt{Prospector} model spectrum for A, the plausible host of FRB 20200223B, plotted along with the flux density in different wavelength bands.  \emph{Right:} Best-fit \texttt{Prospector} model spectrum for B, the plausible host of FRB 20190110C. Different wavelength broadband filters are represented in different colors from ultraviolet to infrared from GALEX, SDSS, 2MASS, SPITZER, and WISE catalogs. The best-fit model spectra are used to estimate different physical properties for A and B, as summarized in Table \ref{tab: A-galaxy-properties}.}
    \label{fig:R49prospector-spectrum}
\end{figure*}

The best-fit SED profile of A (left panel) and B (right panel) are shown in Figure \ref{fig:R49prospector-spectrum}. Since we did not fit the spectra of galaxy B, we plot only the photometry and the fitted model. In Table~\ref{tab: A-galaxy-properties} we present values for stellar mass, age, and SFR from our \texttt{Prospector} analysis of galaxies A and B along with values for stellar mass and SFR for galaxy C from \citep{Chang2015}.  \texttt{Prospector} estimates SFR using the posteriors for the stellar mass, galaxy age, and decay time scale. Based on these results, we find that galaxies A, B and C are all relatively massive with M$_{\rm{gal}} \approx 2-6 \times 10^{10}$ M$_{\odot}$. For galaxies B and C we combine these masses with SFRs estimated from the spectroscopic analysis described above to compute specific star formation rates (see Table~\ref{tab: A-galaxy-properties}).

\begin{deluxetable}{llccc}
\tabletypesize{\small}
\tablecaption{Photometric data used for \texttt{\texttt{Prospector}} SED fitting for Galaxies, A and B \label{tab:prospector-sed}}
\tablehead{\colhead{Dataset} & \colhead{Filter} & \colhead{Central } & \colhead{A Flux density}  & \colhead{B Flux density} \\
\colhead{} & \colhead{} & \colhead{Wavelength (\AA)} & \colhead{}  & \colhead{} \\}
\startdata
     GALEX & FUV & 1528 & 1.68$\times$10$^{-8}$ & 4.85$\times$10$^{-9}$ \\
           & NUV & 2271 & 2.23$\times$10$^{-8}$ & 6.98$\times$10$^{-9}$ \\
     SDSS & u & 3543 & 6.83$\times$10$^{-8}$ & 1.23$\times$10$^{-8}$ \\
          & g & 4770 & 2.18$\times$10$^{-7}$ & 3.49$\times$10$^{-8}$ \\
          & r & 6231 & 3.72$\times$10$^{-7}$ & 5.54$\times$10$^{-8}$ \\
          & i & 7625 & 4.95$\times$10$^{-7}$ & 7.43$\times$10$^{-8}$ \\
          & z & 9134 & 6.38$\times$10$^{-7}$ & 7.91$\times$10$^{-8}$ \\
   2MASS & J & 12350  & 9.28$\times$10$^{-7}$ & - \\
         & K & 16620 & 5.95$\times$10$^{-7}$ & - \\
         & Ks & 21590 & 7.58$\times$10$^{-7}$ & - \\
   WISE & W1 & 33526 & 2.19$\times$10$^{-7}$ & 4.32$\times$10$^{-8}$ \\
         & W2 & 46028 & 1.49$\times$10$^{-7}$ & 2.79$\times$10$^{-8}$ \\
   Spitzer$^{a}$ & 3.6$\mu$m & 35510 & - & 2.71$\times$10$^{-8}$ \\
           & 4.5$\mu$m & 44930 & - & 1.98$\times$10$^{-8}$ \\
           & 5.8$\mu$m & 57300 & - & 1.86$\times$10$^{-8}$ \\
           & 8.0$\mu$m & 78730 & - & 1.24$\times$10$^{-8}$ \\
\enddata
Note:  All flux densities are in maggies. \\
$^{a}$ Flux densities are for 3.8\,arcsecond diameter aperture. 

\end{deluxetable}

\begin{deluxetable*}{llccc}
\tabletypesize{\small}
\tablecaption{Priors used for \texttt{\texttt{Prospector}} SED fitting for Galaxies, A and B \label{tab:prospector-prior}}
\tablehead{\colhead{Prior description} & \colhead{} & \colhead{} & \colhead{A} & \colhead{B} 
}
\startdata
     Total stellar mass formed & log(M$_{\mathrm{*}}$/M$_{\mathrm{\odot}}$) & log-uniform &  min$=$9, max$=$11 & min$=$9, max$=$12 \\
     Stellar metallicity & log(Z/Z$_{\mathrm{\odot}}$) & top-hat &  min$=$ $-$2, max$=$1 & min$=$ $-$5, max$=$ 1 \\
Diffuse V-band dust optical depth & dust2 & top-hat & min$=$0, max$=$0.3 & min$=$0, max$=$5 \\
Stellar population age of the galaxy & t$_{\mathrm{age}}$ (Gyrs) & log-uniform & min$=$5, max$=$12 & min$=$1, max$=$10 \\
E-folding time of the SFH & $\tau$ (Gyrs) & log-uniform & min$=$0.05, max$=$10 & min$=$0.05, max$=$10 \\
\enddata
\end{deluxetable*}

\subsection{Constraints on the presence of persistent radio sources} \label{subsec:prs-constraint}
In the field of FRB 20200223B, we found a source in the high-resolution LOFAR Two-metre Sky Survey \cite[LoTSS;][]{Shimwell2022} LoTSS catalog, ILT J003304.67+284952.4 with an integrated flux density of 2.96$\pm$0.59\,mJy and an isotropic radio luminosity of 2.7(5) $\times$ 10$^{29}$\,erg\,s$^{-1}$\,Hz$^{-1}$ at 144\,MHz given the luminosity distance of 277.81\,Mpc. The radio source is resolved and extended in the LoTSS, with a full-width half maximum (FWHM) of the major axis width of 19.53" (LoTSS beam size, $\theta = 6"$) and also spatially coincident with galaxy A. The radio contours match the optical morphology implying that the radio emission originates from star-formation activity from the galaxy (see the left panel of Figure ~\ref{fig:R49sdss-C64sdss-image}). We used the VLASS upper limit of $<$0.69\,mJy (5$\sigma$) at 3\,GHz and the LoTSS detection at 144\,MHz to estimate an upper limit on the spectral index (S$_{\mathrm{\nu}} \approx \nu^{\alpha}$) of the radio source as $\alpha_{\mathrm{radio}} < -0.48\pm0.09$ which is consistent with star formation in the galaxy. Also, the radio-to-optical flux ratio of $<$ 0.4 is consistent with star formation as opposed to AGN activity (radio-to-optical flux ratio, $F_\mathrm{1.4\,GHz}/F_\mathrm{r-mag}$ $>$ 1.4) \cite[e.g.,][]{mc99, vcb+08,pmt+09}. Therefore, we conclude that the radio source is likely a low-frequency emission coming from star formation in the host galaxy. Due to the large beam size, $\theta = 6"$ of LoTSS, we cannot rule out a faint PRS embedded in this emission.

In the field of FRB 20190110C, we found unresolved radio sources in FIRST (FIRST J163717.8$+$412634), VLASS (VLASS 1QLCIR J163717.83$+$412633.9: Epoch 1 and 2), and LoTSS (ILT J163717.72$+$412633.8) catalogs in the field spatially coincident with one another but not with the plausible host galaxy, B. There is no source found in NVSS down to a 5$\sigma$ threshold of 12.5 mJy/beam. The integrated radio flux density for the FIRST source is 0.91$\pm$0.15\,mJy; for the first epoch of VLASS is 0.85$\pm$0.11\,mJy; for the second epoch of VLASS is 0.72$\pm$0.11\,mJy and for the LoTSS source is 8.67$\pm$0.14\,mJy. The detailed analysis of this radio source will be published by Ibik et al. (in prep.). We report the measured upper limits from the B location in Table ~\ref{tab:radio-catalogue details}. 

In the field of FRB 20191106C, we found five LoTSS sources with one (ILT J131819.22+425958.9, shown in the right panel of Figure~\ref{fig:R49sdss-C64sdss-image}) spatially coincident with the plausible host galaxy, C. For simplicity, we show only the radio source contours that are spatially coincident with the plausible host galaxy.  This source, ILT J131819.22+425958.9, has an integrated flux density of 3.29$\pm$0.14\,mJy and a radio luminosity of 1.05$\pm$0.04 $\times$ 10$^{30}$\,erg\,s$^{-1}$\,Hz$^{-1}$ at 144\,MHz given the luminosity distance of 516.60\,Mpc. The radio source is slightly resolved in the LoTSS, with an FWHM major axis width of 7.76", and is offset by 4.0$\pm$1.2\,kpc from the centre of the host galaxy. We used the VLASS upper limit of $<$0.63\,mJy (5$\sigma$) at 3\,GHz and the LoTSS detection at 144\,MHz to estimate a spectral index $\alpha_{\mathrm{radio}} < -0.54\pm0.02$. The spectral index and the radio-to-optical ratio of $<$ 0.9 are both consistent with star formation as opposed to AGN activity. Therefore, we conclude that the radio source is likely a low-frequency emission coming from star formation in the host galaxy. Due to the large beam size, $\theta = 6"$ of LoTSS, we cannot rule out a faint PRS if embedded in the emission.

\subsection{Estimates of Host Dispersion Measures} \label{subsec:implication-DM-host}
To characterize DM$_\mathrm{host}$ of a galaxy, we first need to properly assess any additional foreground contributions to the DM from intervening structures. Typically, using the equation, DM$_\mathrm{host}$ $=$ DM$_\mathrm{observed}$ $-$ DM$_\mathrm{MW}$(NE2001) $-$ DM$_\mathrm{halo}$ $-$ DM$_\mathrm{IGM}$, any missed or unaccounted intervening media will reduce the estimated value of DM$_\mathrm{host}$. We searched within our FRB baseband positions for foreground galaxy clusters using data from \citet{Wen2009,Wen2012,Wen2015,Wen2018,Banerjee2018}, Galactic \ion{H}{2}-regions using the H$_{\mathrm{\alpha}}$ maps from \citet{green2019revised,Abb+14}, Galactic star-forming regions using CO maps from \citet{ave02,Rice2016}, nearby galaxies from GLADE v2.3 \citep{dgd+18} and dwarf galaxies from \cite{mcc12}. 

The results show that there are no star-forming or \ion{H}{2}-regions in the foreground of the searched FRB repeaters in our sample listed in Table \ref{tab:catalogue details}. We will now estimate DM$_\mathrm{host}$ for the 3 FRBs with plausible hosts discussed in this work and discuss the result of the few cases where there is at least one galaxy cluster or a galaxy in the region.

 For the case of FRB\,20200223B, in the absence of any object along the line-of-sight of the plausible host, we use $z_\mathrm{spec} = 0.06024(2)$ to obtain DM$_\mathrm{IGM}$ = 57.12(2)\,pc\,cm$^{-3}$ following Equation 2 of \cite{Macquart_2020}. We chose a range of DM$_\mathrm{halo}$ with the lower bound and a higher bound that will produce a physical DM$_\mathrm{host}$ for each FRB but still within the given range reported by \citet{Dolag2015,Yamasaki2020,Cook2023} which is 30 $-$ 98\,pc\,cm$^{-3}$ for this FRB. Taking note of the DM$_\mathrm{MW}$(NE2001) $=$ 45(9)\,pc\,cm$^{-3}$, we estimate DM$_\mathrm{host} = 1$ - $73 $\,pc\,cm$^{-3}$.

We do not find any object along the line-of-sight of the host galaxy of FRB 20190110C. We therefore use z$_\mathrm{spec} = 0.12244(6)$ for the host galaxy to directly estimate DM$_\mathrm{IGM}$ and hence DM$_\mathrm{host}$. For DM$_\mathrm{MW}$(NE2001) $=$ 37(7)\,pc\,cm$^{-3}$ and assuming DM$_\mathrm{halo}$ = 30 $-$ 65\,pc\,cm$^{-3}$, we obtain DM$_\mathrm{IGM}$ = 117.75(6)\,pc\,cm$^{-3}$ and DM$_\mathrm{host} = 2$ - $33$\,pc\,cm$^{-3}$.

In the field of FRB 20191106C, there is a galaxy cluster, WHL J131827.1+425803 \citep{Wen2015} with z$_\mathrm{spec} = 0.47$. Given $z_\mathrm{max} = 0.36$, we do not expect that the galaxy cluster will affect the DM of the FRB. There are many other galaxies in the field but none of them has a redshift that could contribute to the DM of the FRB. In the absence of any intervening medium, we estimate DM$_\mathrm{host} =$ 103 $-$ 187\,pc\,cm$^{-3}$ from DM$_\mathrm{IGM}$ = 103.29(1)\,pc\,cm$^{-3}$ at z$_\mathrm{spec} = 0.10775(1)$ of the host galaxy, assuming DM$_\mathrm{halo}$ = 30 $-$ 111\,pc\,cm$^{-3}$ and DM$_\mathrm{MW}$(NE2001) $=$ 25(5)\,pc\,cm$^{-3}$.

For the rest of this section, we take note of any possible intervening structures for the other 10 repeaters in our sample, for which we do not yet have host associations, but for which one should beware if host galaxy associations are made in the future.
We found a nearby galaxy cluster (WHL J172540.4+550250) \citep{Wen2009} with redshift $z = 0.43$ in the field of FRB 20190804E. However, we do not think that this galaxy cluster could affect the DM of the FRB as it has $z_\mathrm{max} = 0.37$.
In the field of FRB 20201114A with $z_\mathrm{max} = 0.33$, we found a GLADE galaxy (HyperLEDA catalog number: 2745781) with z$_\mathrm{spec} = 0.08$ \citep{dgd+18} and 4 other galaxies from the DESI catalog, all having $z < z_\mathrm{max}$. 

We also found 4 galaxy clusters, namely WHL J010806.9+182752 (z$_\mathrm{spec} = 0.17$), WHL J010830.7+183100 (z$_\mathrm{spec} = 0.56$), WHL J010833.0+182514 (z$_\mathrm{spec} = 0.46$) \citep{Wen2015}, and AMF9 J010830.7+182754 (z$_\mathrm{spec} = 0.19$) \citep{Banerjee2018} in the field of FRB 20200929C. Since their redshifts are lower compared to the FRB's $z_\mathrm{max} = 0.44$, it is plausible that WHL J010806.9+182752 and AMF9 J010830.7+182754 could affect the DM of the FRB depending on the location and redshift of the host galaxy. 

\subsection{Comparison to Previous FRB Hosts} \label{subsec:comparison-to-other-FRBs}

\begin{figure*}
 \centering
\includegraphics[width=0.95\textwidth,height=0.45\textheight]{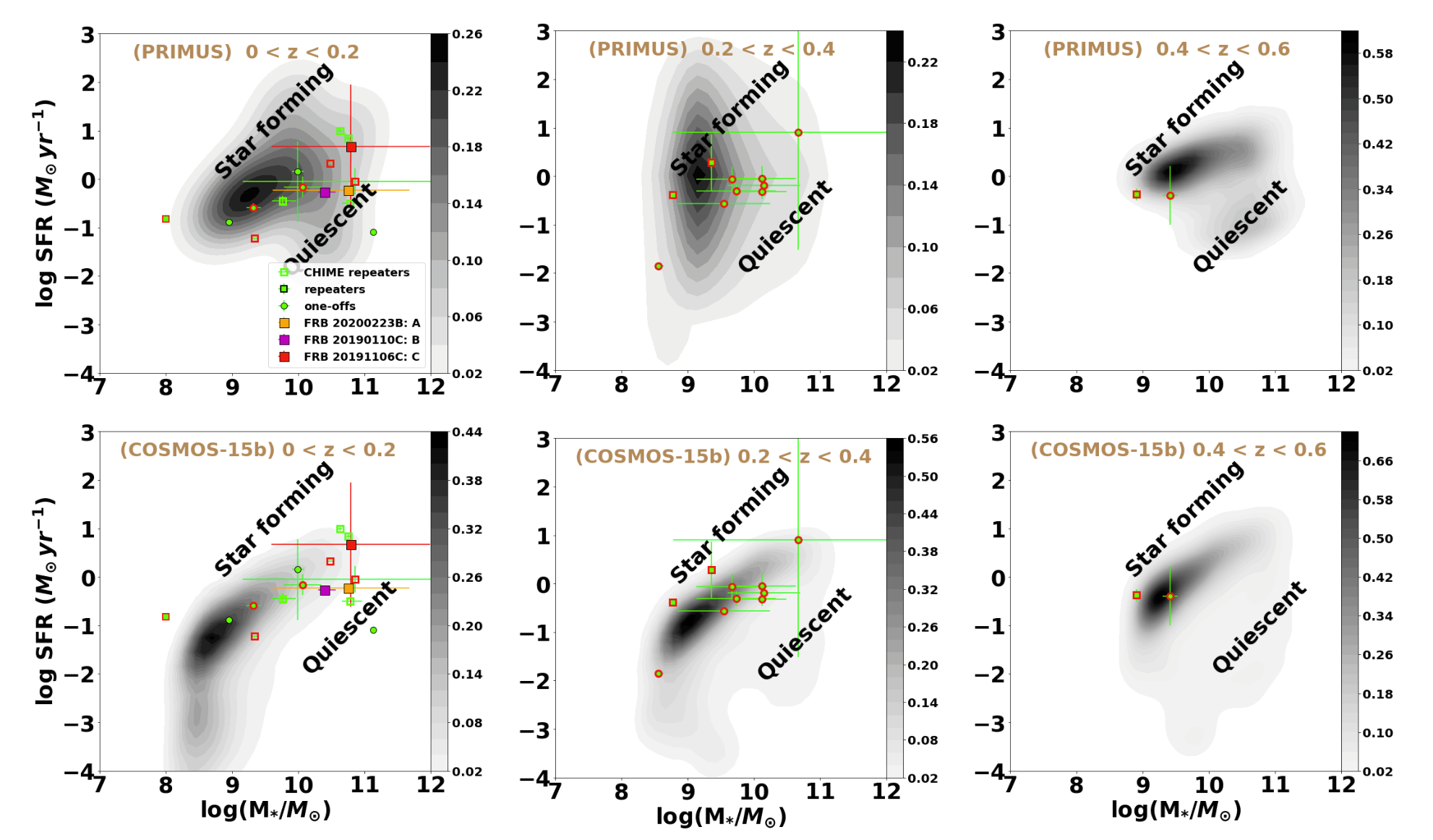}
 \caption{Star formation rate versus stellar mass distribution of FRB hosts in three redshift ranges. The squares and circles in each plot represent repeating and one-off FRBs respectively \citep{Mahony2018, Heintz2020, Mannings2021, Fong2021, Bhardwaj_2021, Bhandari2021, BhardwajR42021, Niu2021, Bhandari2022, Michilli2022, Sharma2023}) while the ones with robust host associations ($\sim <$1 arcsecond) have red edge color.
 The hollow green squares are other CHIME/FRB repeaters while the filled red, purple, and orange squares indicate the 3 host galaxies proposed here for CHIME/FRB repeaters. The grey background is a kernel density distribution of the PRIMUS survey galaxy sample for the top panel and COSMOS-15b for the bottom panel. While the hosts of repeating FRBs show diverse SFR behavior, only a few FRBs are associated with transitioning and quiescent galaxies. Most FRB hosts are slightly offset below the star-forming main sequence when compared to the PRIMUS dataset in the redshift range 0 $<$ z $<$ 0.2. While this effect is less strong compared to COSMIS-15b data, in each case, a subset of FRB-repeater hosts (including those presented here) still exhibit low star formation for their galaxy mass. We discuss this issue further in \S \ref{subsec:comparison-to-other-FRBs}.} \label{fig:comp-sfr}
 \end{figure*}

\begin{figure*}
    \centering   \includegraphics[width=0.95\textwidth, height=0.45\textheight]{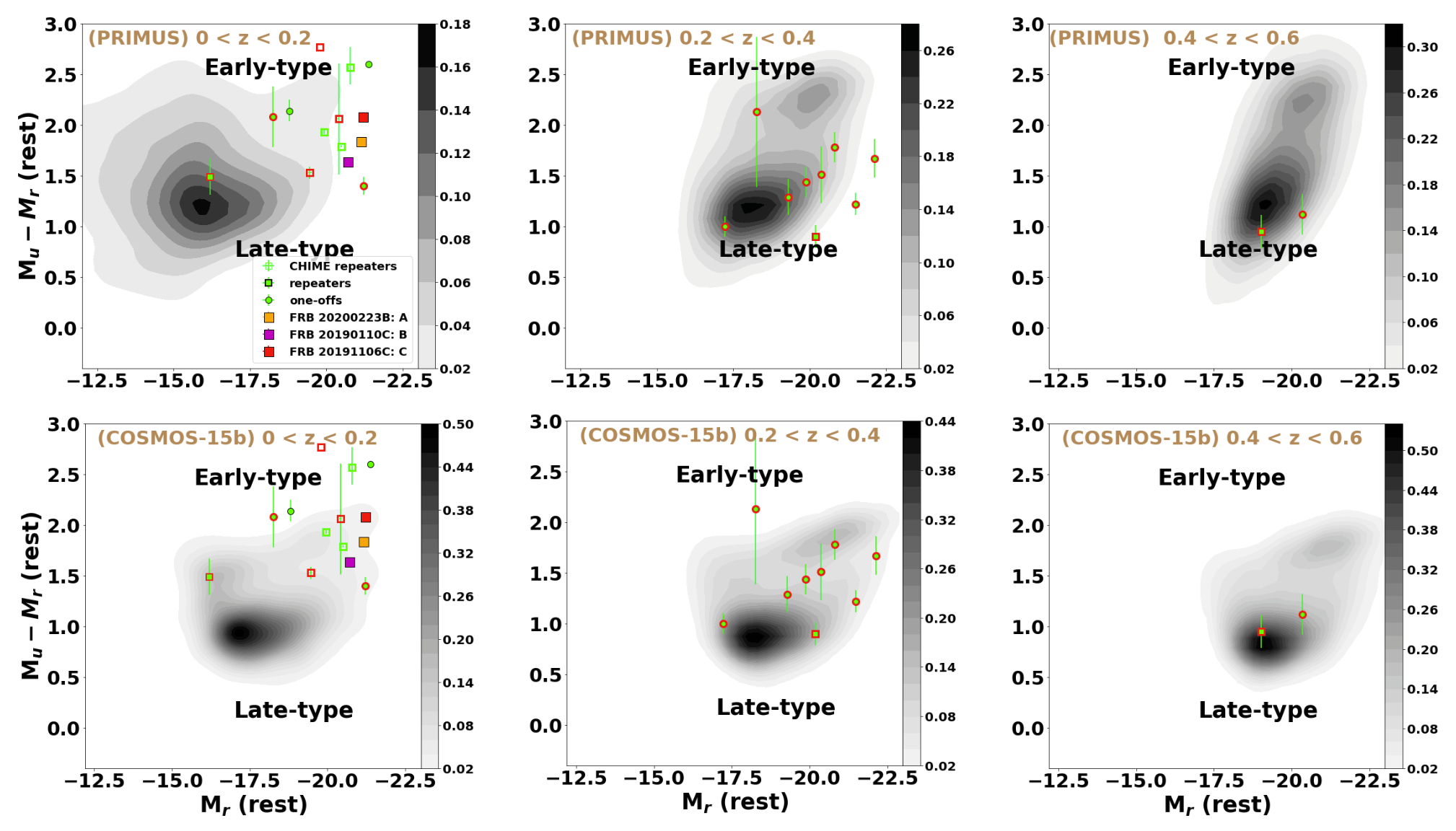}  
    \caption{ Rest-frame color–magnitude diagram in three redshift ranges showing a grey background of the galaxy population from the PRIMUS survey \citep{Moustakas2013} for the top panel and the field galaxies from the COSMOS-15 survey \citep{Leja2020} for the bottom panel. The squares and circles in each plot represent repeating and one-off FRBs respectively \citep{Mahony2018, Heintz2020, Mannings2021, Fong2021, Bhandari2021, Bhardwaj_2021, BhardwajR42021, Bhandari2022, Michilli2022, Sharma2023} while the ones with robust host associations ($\sim <$1 arcsecond) have red edge color.
    The hollow green squares are other CHIME/FRB repeaters while the filled red, purple, and orange squares indicate the 3 host galaxies proposed here for CHIME/FRB repeaters. Many FRBs come from bright galaxies and there is a diversity of early-type, late-type, and ``green-valley'' (parameter space between the two peaks--- early-type and late-type) galaxies among repeaters and one-offs. Approximately half of the FRB-repeater hosts show evidence of being in the transitional green-valley and early-type irrespective of datasets. See \S \ref{subsec:comparison-to-other-FRBs} for details.}
    \label{fig:comp-color}
\end{figure*}

\begin{figure}
\centering
\includegraphics[width=0.5\textwidth,height=0.35\textheight]{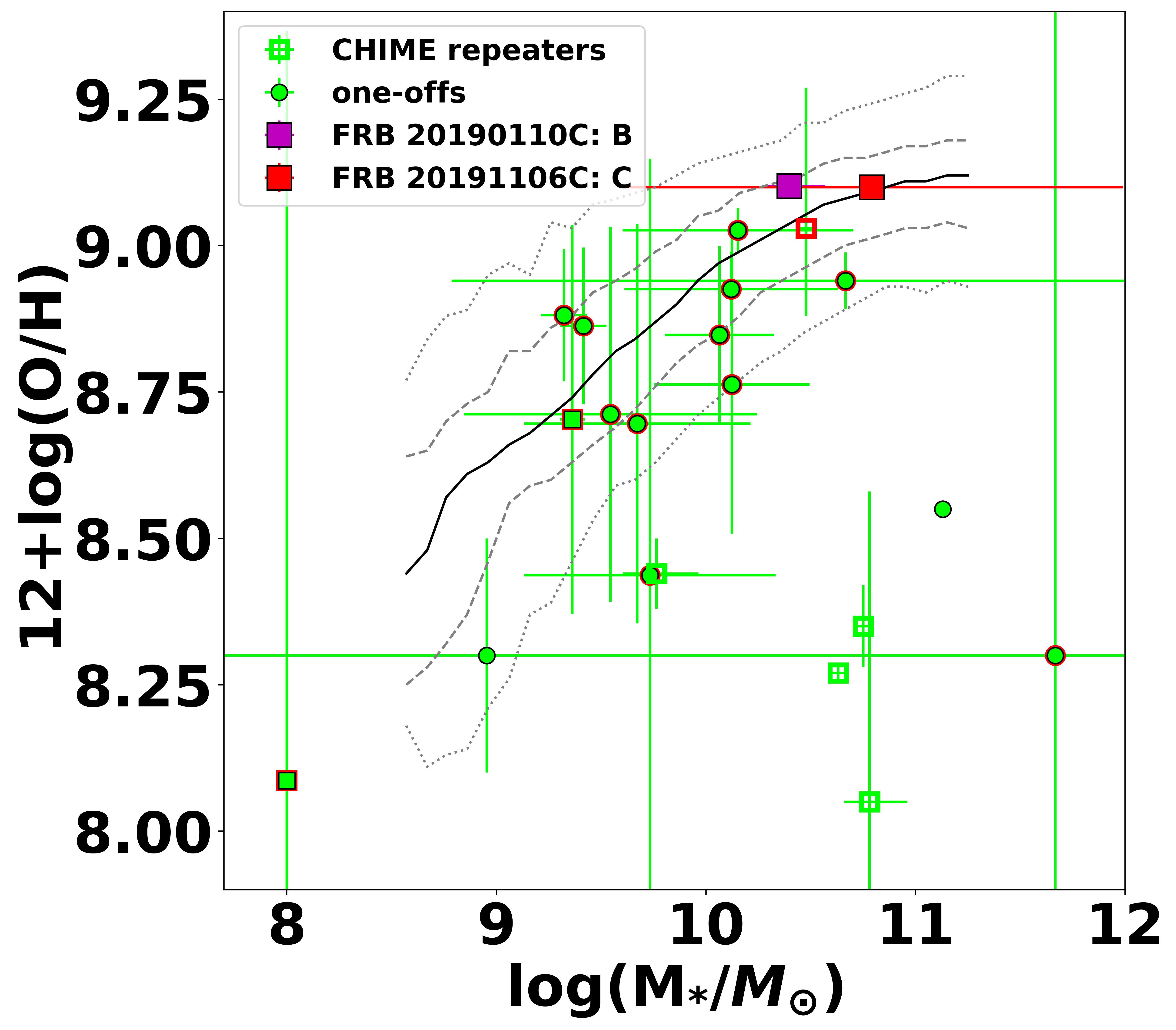}
 \caption{A plot of metallicity versus the stellar mass of galaxies. FRB repeaters and one-offs are represented with squares and circles respectively \citep{Heintz2020, Mannings2021, BhardwajR42021, Bhandari2021, Michilli2022, Sharma2023} while the ones with robust host associations ($\sim <$1 arcsecond) have red edge color. The hollow green squares are other CHIME/FRB repeaters while the filled red and purple squares indicate 2 out of the 3 host galaxies proposed here for CHIME/FRB repeaters. The third galaxy, A was not included in this plot since the nebular emission is contaminated by AGN activity. The solid black line is the median of the SDSS metallicity-stellar mass distribution, and the grey dashed and dotted lines cover the 68\% and 95\% confidence regions \citep{Tremonti2004}. Overall, there is no metallicity preference between repeaters and one-offs but most of them are slightly below the solid line. See \S \ref{subsec:comparison-to-other-FRBs} for details.} \label{fig:comp-met}
 \end{figure}

Based on the results found above, the three probable FRB host galaxies presented here are all relatively massive galaxies with signatures of ongoing star formation and roughly solar metallicity. Here, we compare these properties to those of other known FRB hosts.  To aid this comparison, we plot galaxies A, B, and C along with other FRB hosts on a BPT diagram (Figure~\ref{fig:BPT-plot}), a mass-star formation rate diagram (Figure~\ref{fig:comp-sfr}), a color-magnitude diagram (Figure~\ref{fig:comp-color}) and a mass-metallicity diagram (Figure~\ref{fig:comp-met}). In all plots, FRB hosts from the literature are shown in green, with repeaters designated as squares and one-offs as circles. Previous repeating FRBs that were identified by CHIME/FRB are shown as open squares. FRBs with robust host association are shown as red-edge-colored markers.

\emph{BPT diagram: } First, to compare the dominant sources of photoionization in FRBs host galaxies, we add FRBs from the literature to the BPT diagram described in \S \ref{subsec:emission-line} to get Figure \ref{fig:BPT-plot}. A total of 15 FRB hosts' emission line fluxes were taken from \cite{Heintz2020}, \cite{Mannings2021}, \cite{Fong2021}, \cite{BhardwajR42021}, \cite{Bhandari2021}, and \cite{Michilli2022} and shown in green squares and circles for 5 repeating (the host of FRB 20190303A are two merging galaxies which leads to 6 open squares in the Figure) and 10 non-repeating FRBs. Out of the 15 FRBs, 3 repeaters and 9 one-offs have robust (FRBs with localization region of $\sim <$1 arcsecond) host associations (see markers with red edge colors in Fig ~\ref{fig:BPT-plot}).
The plot shows that the 3 host galaxies' (A, B, C) line ratios are broadly consistent with those of most FRB hosts, although galaxy A would be the first host for a \emph{repeating} FRB with signatures of AGN activity.  Galaxies B and C, support the trend identified by \cite{Bhandari2021} that most repeating FRBs happen in galaxies where star formation is the dominant source of photo-ionization, thus disfavoring progenitor channels where AGN activity is required \citep[e.g.][]{Katz2017, Vieyro2017, Gupta2018}.

\emph{SFR-Mass: } In Figure~\ref{fig:comp-sfr}, we compare the rate of star formation versus the stellar mass of FRB host galaxies in three redshift bins, 0 $<$ z $<$ 0.2, 0.2 $<$ z $<$ 0.4, and 0.4 $<$ z $<$ 0.6.  SFRs and masses for 24 FRB hosts were extracted from \cite{Mahony2018}, \cite{Heintz2020}, \cite{Mannings2021}, \cite{Fong2021}, \cite{Bhardwaj_2021}, \cite{Bhandari2021}, \citep{BhardwajR42021}, \cite{Niu2021}, \cite{Bhandari2022}, \cite{Michilli2022}, and \cite{Sharma2023} and shown in green squares and circles for 10 repeating (the host of FRB 20190303A are two merging galaxies which leads to 11 open squares in the Figure) and 14 non-repeating FRBs.  Out of the 24 FRBs, 7 repeaters and 11 one-offs have robust (FRBs with localization region of $\sim <$1 arcsecond) host associations (see markers with red edge colors in Fig ~\ref{fig:comp-sfr}).
We plot the location of galaxies A, B, and C using SFRs, and stellar masses measured from \texttt{Prospector} SED fitting described above. 

In the top left panel, kernel density contours represent galaxies from the PRism MUlti-object Survey (PRIMUS) survey \citep{Moustakas2013} with redshifts z $<$ 0.5. In the bottom left panel, kernel density contours represent galaxies from the COSMOS-15 sample that were modeled by \citet{Leja2020}, referred to as COSMOS-15b. \citet{Bhandari2021} previously identified that most FRB host galaxies are offset slightly below the star-forming main sequence, showing low star formation rates for their masses when compared to the PRIMUS dataset.  This offset is visible in our figure (see Figure~\ref{fig:comp-sfr}), where the star-forming main sequence is shifted about half an order of magnitude lower in the upper panels (PRIMUS) of Figure~\ref{fig:comp-sfr}. In agreement with the result of \cite{Gordon2023}, when compared to the lower panel (COSMOS15b), the galaxies mostly fall within the star-forming main sequence in closer alignment to many of the known FRB hosts. 
While reconciling this discrepancy is beyond the scope of this work, we note the following: \\
i) Our datasets are a mix of both spectroscopic measurements and SED-derived information and so some are model-dependent. This makes it difficult to select direct analogous survey data for comparison;
ii) We note that below redshift $<$ 0.2, the PRIMUS sample has fewer galaxies compared to higher redshift field galaxies and lacks quality SFR and stellar mass measurements within this redshift range; 
iii) The PRIMUS dataset redshifts are however measured from spectroscopy and the derived galaxy property values are from parametrized star formation history; 
iv) The COSMOS-15b dataset on the other hand has redshift measurements from a mixture of spectroscopic and photometric data while its galaxy properties are derived values from a non-parametric continuity SFH.

Compared to \emph{both} comparison datasets, galaxies A, B, and C (and many of the other CHIME repeaters) lie along and below the star-forming main sequence. In addition, the WISE plot location of galaxy A, showing ongoing star formation coupled with emission line signatures showing AGN activity, is also consistent with a galaxy in a transition phase. Our findings show that most FRBs appear to come from star-forming galaxies across the FRB redshift range observed so far. While the hosts of repeating FRBs show diverse SFR behavior, only a few FRBs are associated with transitioning and quiescent galaxies. Most FRB hosts are slightly offset below the star-forming main sequence when compared to the PRIMUS dataset in the redshift range 0 $<$ z $<$ 0.2. While this effect is less strong compared to the COSMOS-15b data, in each case, a subset of FRB-repeater hosts (including those presented here) still exhibit low star formation for their galaxy mass.

Overall, even though most FRB hosts prefer star-forming hosts, there may be an observational bias against finding FRBs in star-forming regions due to large possible levels of pulse dispersion and scattering in such regions. This could lead to decreased sensitivity to repeating bursts \citep{Gordon2023}. The difference in the SED-modelling methods used for the two field galaxy datasets as well as insufficient data for a given redshift range are possible factors that could affect the interpretation of this result.

\cite{Gordon2023} also used the mass-doubling number criterion (the number of times the stellar mass doubles within the age of the universe at redshift at the redshift of the galaxy assuming a constant specific star formation rate, sSFR) described by \cite{Tacchella2022} to classify galaxies as star-forming, transitioning and quiescent. We adopted this method for all the galaxies in our comparison sample as well as galaxies A, B, and C (even though we caution that the criterion was developed for galaxy properties derived using non-parametric star formation histories). Our results of the literature comparison sample show that most FRB hosts are star-forming, few are transitioning and one is quiescent. In addition, Galaxies A and B are both transitioning while C is star-forming according to the criteria with mass-doubling numbers of 0.1363, 0.2686, and 0.9547 respectively. Together, we note that all four galaxies classified as transitioning by this metric were repeaters. The overall outcome (that many FRB hosts are star-forming) is however consistent with the conclusions of \cite{Gordon2023} and \cite{Sharma2023}.

\emph{Color-Magnitude: } To further examine the possibility that a subset of repeating FRB hosts are in a transitional phase, in Figure \ref{fig:comp-color}, we compare the color–magnitude properties of the FRB hosts in three redshift bins, 0 $<$ z $<$ 0.2, 0.2 $<$ z $<$ 0.4 and 0.4 $<$ z $<$ 0.6. This can be a useful indicator of the class of stellar population in the galaxies, which is less model-dependent. Specifically, the plot of the rest-frame M$_{\mathrm{u}}-M_{\mathrm{r}}$ colors versus their absolute r-band magnitudes, M$_{\mathrm{r}}$ is overplotted with the kernel density contours of field galaxies. The top panel again compares to galaxies from the PRIMUS survey with redshifts z $<$ 0.5 and the lower panel with galaxies of the COSMOS-15b \citep{Leja2020} dataset retrieved from the parent catalog \citep{Laigle2016cosmos}. A total of 21 FRB hosts' color-magnitude values were extracted from \cite{Mahony2018}, \cite{Heintz2020}, \cite{Mannings2021}, \cite{Fong2021}, \cite{Bhandari2021}, \cite{Bhardwaj_2021}, \cite{BhardwajR42021}, \cite{Bhandari2022}, \cite{Michilli2022}, and \cite{Sharma2023} and shown in green squares and circles for 8 repeating (the host of FRB 20190303A are two merging galaxies which leads to 9 open squares in the Figure) and 13 non-repeating FRBs. Out of the 21 FRBs, 6 repeaters and 11 one-offs have robust (FRBs with localization region of $\sim <$1 arcsecond) host associations (see markers with red edge colors in Fig ~\ref{fig:comp-color}).
Galaxies A, B, and C are bright but fall within the overall range of properties seen in the literature for the FRB hosts. Many FRBs come from bright galaxies and there is a diversity of early-type, late-type, and ``green-valley'' (parameter space between the two peaks--- early-type and late-type) galaxies among repeaters and one-offs. 

 Our result shows that the observed colors for all three probable hosts presented in this manuscript fall within the ``green valley'' when compared to the PRIMUS dataset, which would be consistent with having low star formation rates for their stellar masses (We note that the PRIMUS dataset has very few quiescent galaxies in the redshift range 0 $<$ z $<$ 0.2 which is why there is no corresponding ``early-type'' island in the left top panel).  In particular, with the addition of these hosts, \emph{half} of the FRB repeaters shown in Figure \ref{fig:comp-color} have colors between 1.6\,mag $<$ M$_{\mathrm{u}}-M_{\mathrm{r}}$ $<$ 2.1 mag.

When comparing to the COSMOS-15b dataset, we again see that the three probable hosts described here show redder colors than a majority of late-time galaxies. (We note that the \citealt{Leja2020} sample has very few quiescent galaxies, which is why there is a small corresponding ``early-type'' islands in all the lower panels). We note that this is a slightly different behaviour that was observed with the SFR-Stellar mass plot for the COSMOS-15b dataset (lower panels Figure~\ref{fig:comp-sfr}) where many FRBs fell directly on the SFR main sequence. Given that the exact same set of galaxies are plotted in both figures, this distinction must stem from methods used to measure star-formation rate and stellar mass for the COSMOS-15b and FRB datasets (see above).

Approximately half of the FRB-repeater hosts show evidence of being in the transitional green-valley and early-type, irrespective of datasets. The moderate and low SFR seen for most FRB hosts are consistent with the location on the color-magnitude diagram. Most of the hosts are in the green valley and have a redder color possibly because of their low current star formation. This result is however non-intuitive as previous analysis shows that FRB hosts are mostly star-forming (which is mostly associated with late-type galaxies). It is possible that many FRBs are coming from star-forming regions of either late-type or green-valley galaxies. Indeed, we see evidence above for ongoing star formation within the potential FRB hosts presented here. Alternatively, this may be evidence for an additional, older, FRB progenitor channel, which is slightly overrepresented in the repeater population. Indeed, there has been recent evidence for some FRBs coming from older environments, such as  FRB 20200120E found in a globular cluster of a spiral galaxy \citep{Bhardwaj_2021, kirsten2021}. Finally, it is possible that this outcome is a selection bias since it may be easier to see FRBs from early-type galaxies given that they are not obscured by a high degree of dispersion and scattering from ionised gas in the hosts \citep{Mannings2021}.

\emph{Mass-Metallicity: } Finally, Figure \ref{fig:comp-met}
shows the distribution of host galaxies' metallicities and stellar masses.  These are shown in comparison to the \citet{Tremonti2004} stellar mass versus gas-phase metallicity relationship found from SDSS data of redshifts z $\sim$ 0.1. The solid black line is the median, and the grey dashed and dotted lines cover the 68\% and 95\% confidence regions.  Host information for 19 other FRBs was extracted from \cite{Heintz2020, Mannings2021, BhardwajR42021, Bhandari2021, Michilli2022, Sharma2023} consisting of 6 repeaters (the host of FRB 20190303A are two merging galaxies which leads to 7 open squares in the Figure) and 13 one-offs. Out of the 19 FRBs, 3 repeaters and 11 one-offs have robust (FRBs with localization region of $\sim <$1 arcsecond) host associations (see markers with red edge colors in Fig ~\ref{fig:comp-met}). In most cases, we extract raw emission line fluxes and calculate metallicity ourselves using the O3N2 calibration \citep{Hirschauer2018}. In 3 cases where the line fluxes were not available and we could not identify the metallicity calibration used, we took published metallicities directly, acknowledging that a small error may be present in their location within Figure \ref{fig:comp-met} due to systematic offsets between the various strong line metallicity diagnostics \citep{Kewley2008}. Galaxies B and C are both relatively high-mass galaxies that fall directly on the mass-metallicity relationship of  \citet{Tremonti2004}. Overall, there is no metallicity preference between repeaters and one-offs but most of them are slightly below the solid line. This supports the conclusion that FRBs, as a whole, do not require the low metallicity, high specific star-formation rate environments preferred by superluminous supernovae and long-duration gamma-ray bursts \citep{Lunnan2014}.

\emph{Trends with FRB repeat rate: } We used the repeat rates reported by \cite{CHIME2023} for 8 CHIME/FRB events in our sample to compare with host properties (metallicity, SFR, color). Out of these, 7 of them are clustered at similar low repeat rates ($<$1\,hr$^{-1}$) with only one having a high repeat rate ($>$1\,hr$^{-1}$). The results show no relationship between repeat rates and any of the properties given the current data. However, this is a very small dataset; a larger sample collected in the future could provide more insight.

\emph{Trends with FRB DM$_\mathrm{host}$: } We retrieved the FRB DM$_\mathrm{host}$ values of 24 FRBs containing 8 repeaters and 16 one-offs from \cite{Law2022}. We included the DM$_\mathrm{host}$ estimates from FRB\,20200223B, FRB\,20190110C, and FRB 20191106C thereby increasing the sample of repeaters to 11 events. We applied the two-sample Kolmogorov-Smirnov test to the DM$_\mathrm{host}$ histogram distribution for repeaters and non-repeaters. The result is a p-value of 0.4112 ($\sim$41\%). Assuming a null hypothesis at a 95\% confidence interval that the repeaters and non-repeaters have the same local environment, then a p-value that is less than 0.05 (5\%) will reject the null hypothesis. In this case, the p-value is greater than 5\%, thus in agreement with the null hypothesis stating that the two distributions are the same.  We note that there is an outlier among the repeaters which is FRB 190520. However, repeating the test with the outlier removed does not change the conclusion.

In summary, the sample of about 24 known FRB of host galaxies \citep[repeating and non-repeating;][]{Heintz2020, Bhandari2021, Niu2021, Bhandari2022, Michilli2022} covers a broad range of $r$-band absolute magnitude, M$_{\mathrm{u}}$-M$_{\mathrm{r}}$ colors, stellar masses, star formation rates, and metallicities. The three new probable hosts identified here are consistent with this broader population. However, it is notable that through a combination of intermediate M$_{\mathrm{u}}-M_{\mathrm{r}}$ colors (galaxies A, B, C), low star formation rate for a given stellar mass (galaxies A, B, C), mass-doubling timescale (galaxies A, B, C) and/or an elliptical morphology with signatures of ongoing star formation (galaxy C), all three show some evidence of being in a transitional or star-forming regime. Coupled with the previously published hosts, there is tentative evidence that FRB repeaters, in particular, are over-represented in the star-forming and ``green valley'' with many showing M$_{\mathrm{u}}-M_{\mathrm{r}}$ colors redder than the majority of star-forming galaxies.

\section{Summary} \label{sec:summary}

We searched for host galaxy associations for 13 FRB repeaters out of the 25 ``gold'' and 14 ``silver'' samples recently published by \cite{CHIME2023}. The CHIME/FRB baseband localization regions were searched for host galaxies in SDSS and DESI.  
We considered two methods to assess the probability of host association for candidate galaxies within the field. First, we run candidate host galaxies through \texttt{PATH} to estimate their probabilities of association with an FRB. Second, we consider the probability that a galaxy as bright as the host preferred by PATH would be found within the CHIME localization region by chance. When we consider the ``look-elsewhere'' effects associated with the fact that we searched 13 CHIME/FRB localization regions, we do not find evidence for a significant overabundance of bright galaxies. However, under the assumption that FRBs come from galaxies, the Bayesian approach of \texttt{PATH} still allows us to consider the most probable host for an individual FRB.

In the end, we consider two FRBs (FRB\,20200223B, FRB\,20190110C) for which we have galaxy candidates with a maximum association probability that is greater than 90\%. 
We also cautiously discuss a third FRB (FRB 20191106C) with a plausible host galaxy for which the archival surveys have a shallower depth at the maximum redshift of the FRB. We obtained a Max P$_{\mathrm{\texttt{PATH}}}$ $=$ 81.5\% at $\mathrm{P(U)} =0.1$, but this galaxy was strongly preferred over any other galaxy in the field.
The resulting discussed host galaxies for the 3 FRBs (FRB\,20200223B, FRB\,20190110C, and FRB 20191106C) are SDSS J003304.68+284952.6, SDSS 163716.43+412636.2, and SDSS J131819.23+425958.9, respectively.

FRB\,20200223B is likely located in a spiral, star-forming galaxy with evidence for AGN activity. FRB\,20190110C likely comes from a high metallicity, irregular galaxy showing evidence of quenching star formation, while the potential host galaxy of FRB 20191106C is a high metallicity, possibly elliptical with evidence for ongoing star formation. The latter two galaxies follow the standard mass-metallicity relationship observed for nearby galaxies. Moreover, all 3 galaxies are in the transitional phase with regard to star formation and show redder colors than expected for late-type galaxies. This is consistent with the broader repeater host population, which appears to prefer the region between early and late-type galaxies as discussed in \S \ref{subsec:comparison-to-other-FRBs}. In addition, the high metallicity and low specific star-formation rate obtained for these galaxies are different from what is seen for the hosts of superluminous supernovae and long-duration gamma-ray bursts.
However, it is critical to both confirm these host galaxies via arcsecond localization and to continue to increase the number of FRBs with robust host associations, since probabilistic host identification methods are biased towards bright galaxies.

\emph{Facilities:} CHIME, Gemini North \\ 

\emph{Softwares:}  \texttt{FRUITBAT} \citep{Batten2019},
\texttt{PATH} \citep{Aggarwal2021}, \texttt{PYRAF} \citep{pyraf2012}, \texttt{Prospector} \citep{prospector2021}, \texttt{emcee} \citep{Foreman-Mackey2013}, \texttt{python-fsps} \citep{Conroy_2009}, Astropy \citep{astropy2}, Matplotlib \citep{matplotlib}, NumPy \citep{numpy}, SAOImage DS9\citep{ds92003}  \\

\section{Acknowledgement} \label{acknowledgement}
We thank Kasper Heintz and Joel Leja for the supplementary information and code that facilitated this work. We sincerely appreciate Prof. Edward Wright and Chao-Wei Tsai for granting permission to use the WISE diagnostic plot from their paper.

We acknowledge that CHIME is located on the
traditional, ancestral, and unceded territory of the
Syilx/Okanagan people. We are grateful to the staff
of the Dominion Radio Astrophysical Observatory,
which is operated by the National Research Council
of Canada. CHIME is funded by a grant from the
Canada Foundation for Innovation (CFI) 2012 Leading
Edge Fund (Project 31170) and by contributions from
the provinces of British Columbia, Quebec, and Ontario.
The CHIME/FRB Project is funded by a grant from the
CFI 2015 Innovation Fund (Project 33213) and by contributions from the provinces of British Columbia and Quebec, and by the Dunlap Institute for Astronomy and Astrophysics at the University of Toronto. Additional support was provided by the Canadian Institute
for Advanced Research (CIFAR), McGill University, and
the McGill Space Institute thanks to the Trottier Family Foundation and the University of British Columbia. The CHIME/FRB baseband
system is funded in part by a Canada Foundation for
Innovation John R. Evans Leaders Fund award to IHS. 

The Dunlap Institute is funded through an endowment established by the David Dunlap family and the University of Toronto. 

Based on observations obtained at the Gemini Observatory, which is operated by the Association of Universities for Research in Astronomy, Inc., under a cooperative agreement with the NSF on behalf of the Gemini partnership: the National Science Foundation (United
States), the Science and Technology Facilities Council (United Kingdom), the National Research Council (Canada), CONICYT (Chile), the Australian Research Council (Australia), Ministério da Ciência e Tecnologia (Brazil), and SECYT (Argentina). We appreciate the Gemini team for granting us observing time for the program ID: GN-2022B-Q-308.

B.M.G. acknowledges the support of the Natural Sciences and Engineering Research Council of Canada (NSERC) through grant RGPIN-2022-03163, and of the Canada Research Chairs program. MRD acknowledges support from the NSERC through grant RGPIN-2019-06186, the Canada Research Chairs Program, the Canadian Institute for Advanced Research (CIFAR), and the Dunlap Institute at the University of Toronto. V.M.K. holds the Lorne Trottier Chair in Astrophysics \& Cosmology, a Distinguished James McGill Professorship, and receives support from an NSERC Discovery grant (RGPIN-228738-13), and from the FRQNT CRAQ. K.W.M. holds the Adam J. Burgasser Chair in Astrophysics and is supported by an NSF Grant (2008031). A.B.P. is a Banting Fellow, McGill Space Institute (MSI) Fellow, and a Fonds de Recherche du Quebec -- Nature et Technologies (FRQNT) postdoctoral fellow. Z.P. and P. S. are Dunlap Fellows. FRB Research at UBC is funded by an NSERC Discovery Grant and by the Canadian Institute for Advanced Research.  The CHIME/FRB baseband system is funded in part by a CFI John R. Evans Leaders Fund award to IHS. M.B. is a Mcwilliams fellow. F.A.D is supported by the UBC Four Year Fellowship. A.M.C is funded by an NSERC Doctoral Postgraduate Scholarship. K.S. is supported by the NSF Graduate Research Fellowship Program.


\appendix
\section{\texttt{PATH} Probabilities of all candidates}\label{appendix}

Here we present the Tables of probabilities of all the sources in the field of FRB 20200223B, FRB 20190110C, and FRB
20191106C. R$_{50}$ is the half-light radius of the galaxy from the DESI survey, P$_{0.0}$ is the probability that the object is the preferred host galaxy of the FRB given the prior $\mathrm{P(U)} = 0.0$ and P$_{0.1}$ is the probability that the object is the preferred host galaxy of the FRB given the prior $\mathrm{P(U)} = 0.1$.

\begin{deluxetable*}{l|cccc|cc}
\tabletypesize{\small}
\tablecaption{\texttt{PATH} probability for the sources in the field of FRB 20200223B \label{tab: path_A}}
\tablehead{\colhead{Source} & \colhead{RA} &  \colhead{DEC} &  \colhead{R$_{50}$} &  \colhead{r-band} & \colhead{P$_{0.0}$} 
 & \colhead{P$_{0.1}$} \\ 
\colhead{} & \colhead{\degr} &  \colhead{\degr} &  \colhead{"} &  \colhead{mag} & \colhead{} 
 & \colhead{} }
\startdata
1 &	8.269527 & 28.831267 &	5.814174	& 16.080395	& 0.994104 & 0.899497 \\
2 &	8.266982 &	28.829577	& 0.405767	& 22.06333	& 0.001767 & 0.001599 \\
3 &	8.267366	& 28.82594	& 0.646236	& 21.993284	& 0.001427 & 0.001291 \\
4 &	8.263246 &	28.828137 &	0.434656 &	23.971949	& 0.000322 & 0.000291 \\
5 &	8.263912	& 28.828986	& 0.877458	& 23.089138	& 0.000708 & 0.00064 \\
6 &	8.268353	& 28.826364	& 0.439845	& 22.411278	& 0.000992 & 0.000897 \\
7 &	8.271681	& 28.835329	& 0.450015	& 22.615425	&	0.000682 & 0.000617
\enddata
\end{deluxetable*}

\begin{deluxetable*}{l|cccc|cc}
\tabletypesize{\small}
\tablecaption{\texttt{PATH} probability for the sources in the field of FRB 20190110C \label{tab: path_B}}
\tablehead{\colhead{Source} & \colhead{RA} &  \colhead{DEC} &  \colhead{R$_{50}$} &  \colhead{r-band} & \colhead{P$_{0.0}$} 
 & \colhead{P$_{0.1}$} \\ 
\colhead{} & \colhead{\degr} &  \colhead{\degr} &  \colhead{"} &  \colhead{mag} & \colhead{} 
 & \colhead{} }
\startdata
1 & 249.318485	& 41.443412	& 3.379357	& 18.00912	& 0.917943 & 0.779452 \\
2 & 249.324262	& 41.442808	& 0.662556	& 22.32065	& 0.013446 & 0.011418 \\
3 & 249.336069	& 41.448172	& 0.39331	& 21.21785	& 0.025025 & 0.02125 \\
4 & 249.315487	& 41.448435	& 0.337111	& 23.374989	& 0.003773 & 0.003204 \\
5 & 249.317587	& 41.444105	& 2.032882	& 21.900526	& 0.016519 & 0.014027 \\
6 & 249.319692	& 41.437475	& 1.434205	& 23.942047	& 0.002159 & 0.001833 \\
7 & 249.319796	& 41.440578	& 0.571469	& 22.459723	& 0.009711 & 0.008246 \\
8 & 249.32092	& 41.448989	& 0.321152	& 23.473305	& 0.004379 & 0.003719 \\
9 & 249.325318	& 41.441917	& 0.459996	& 23.032875	& 0.007045 & 0.005982 
\enddata
\end{deluxetable*}

\begin{deluxetable*}{l|cccc|cc}
\tabletypesize{\small}
\tablecaption{\texttt{PATH} probability for the sources in the field of FRB 20191106C \label{tab: path_C}}
\tablehead{\colhead{Source} & \colhead{RA} &  \colhead{DEC} &  \colhead{R$_{50}$} &  \colhead{r-band} & \colhead{P$_{0.0}$} 
 & \colhead{P$_{0.1}$} \\ 
\colhead{} & \colhead{\degr} &  \colhead{\degr} &  \colhead{"} &  \colhead{mag} & \colhead{} 
 & \colhead{} }
\startdata
1 & 199.580129	& 42.999713	& \bf{1.601871}	& 17.306004	& \bf{0.951259} & \bf{0.815378} \\
2 & 199.567778 & 43.005932 &	0.361793	& 23.50386	& 0.001473 &	0.001262 \\
3 & 199.568633	& 43.007172	& 1.032621	& 21.765076	& 0.006658	& 0.005707 \\
4 & 199.571617	& 42.998974	& 0.700418	& 22.764788	& 0.00331	& 0.002837 \\
5 & 199.57512	& 42.99899	& 0.467756	& 22.17816	& 0.006046	& 0.005182 \\
6 & 199.575607	& 43.007772	& 0.982723	& 21.274609	& 0.012805	& 0.010976 \\
7 & 199.577994	& 43.000793	& 1.937793	& 23.018389	& 0.003076	& 0.002636 \\
8 & 199.578285	& 42.994688	& 0.400479	& 23.901838	& 0.001067	& 0.000915 \\
9 & 199.580623	& 43.009469	& 1.088066	& 22.449223	& 0.003727	& 0.003195 \\
10 & 199.582184	& 42.998333	& 2.298028	& 21.562593	& 0.009444	& 0.008095 \\
11 & 199.584588	& 43.000609	& 0.749572	& 23.984762	& 0.001136	& 0.000973
\enddata
\end{deluxetable*}

\end{document}